\documentclass[fleqn,11pt]{article}
\usepackage{amscd,amsmath,amssymb,verbatim}
\usepackage{amsthm}
\usepackage{amssymb}
\usepackage{pifont}
\usepackage[pdftex]{graphicx}
\usepackage{epstopdf}
\usepackage[T1]{fontenc}
\usepackage{ae,aecompl}
\usepackage{makecell}
\usepackage{subcaption}
\usepackage{array}
\usepackage{mathrsfs}
\usepackage[nohead]{geometry}
\usepackage[singlespacing]{setspace}
\usepackage{rotating}
\usepackage{multirow}
\usepackage{threeparttable}
\usepackage{enumerate}
\usepackage{placeins}
\usepackage{color, colortbl}
\usepackage{accents}
\usepackage{pdflscape}
\usepackage{longtable}
\usepackage{natbib}
\usepackage{graphicx}
\usepackage{subcaption}
\usepackage{caption}
\usepackage{adjustbox}   
\usepackage{booktabs}    
\usepackage{caption}     

\usepackage{tabularx}
\usepackage{array}
\usepackage{booktabs}
\usepackage{float}

\usepackage{subcaption}
\usepackage{float}
\numberwithin{equation}{section}

\newcommand{\xvec}{\boldsymbol}
\newcommand{\xmat}{\mathbf}

\usepackage{hyperref}

\makeatletter
\newcommand{\distas}[1]{\mathbin{\overset{#1}{\kern\z@\sim}}}%
\newsavebox{\mybox}\newsavebox{\mysim}
\newcommand{\distras}[1]{%
  \savebox{\mybox}{\hbox{\kern3pt$\scriptstyle#1$\kern3pt}}%
  \savebox{\mysim}{\hbox{$\sim$}}%
  \mathbin{\overset{#1}{\kern\z@\resizebox{\wd\mybox}{\ht\mysim}{$\sim$}}}%
}
\makeatother

\renewcommand{\hat}[1]{\widehat{\text{$#1$}}}
\makeatletter
\newsavebox\myboxA
\newsavebox\myboxB
\newlength\mylenA

\renewcommand*\bar[2][0.85]{%
    \sbox{\myboxA}{$\m@th#2$}%
    \setbox\myboxB\null
    \ht\myboxB=\ht\myboxA%
    \dp\myboxB=\dp\myboxA%
    \wd\myboxB=#1\wd\myboxA
    \sbox\myboxB{$\m@th\overline{\copy\myboxB}$}
    \setlength\mylenA{\the\wd\myboxA}
    \addtolength\mylenA{-\the\wd\myboxB}%
    \ifdim\wd\myboxB<\wd\myboxA%
       \rlap{\hskip 0.5\mylenA\usebox\myboxB}{\usebox\myboxA}%
    \else
        \hskip -0.5\mylenA\rlap{\usebox\myboxA}{\hskip 0.5\mylenA\usebox\myboxB}%
    \fi}
\makeatother
\makeatletter
\def\@biblabel#1{\hspace*{-\labelsep}}
\makeatother
\hoffset 
\voffset
\author{Ariane N. Meli Chrisko\thanks{University of Glasgow, UK, email: a.meli-chrisko.1@research.gla.ac.uk} \and Jessie Li \thanks{University of Glasgow, UK, email: 
jessiewingli1@gmail.com}  \and Philipp Otto\thanks{University of Glasgow, UK, email: philipp.otto@glasgow.ac.uk} \and Wolfgang Schmid\thanks{European University Viadrina, Frankfurt (Oder), Germany, email: schmid@europa-uni.de}  }\medskip

\title{Comparative Analysis of Spatiotemporal Volatility Models: An Empirical Study on Financial Network Series}

\begin{document}
\maketitle
\sloppy

\singlespacing

\begin{abstract}
\noindent
Various spatiotemporal and network GARCH models have recently been proposed to capture volatility interactions, such as the transmission of market risk across financial networks. These approaches rely heavily on the specification of the adjacency or spatiotemporal weight matrix, for which several alternatives exist in the literature.  
This paper evaluates the out-of-sample forecasting performance of a range of spatiotemporal volatility models and multivariate GARCH benchmarks under nine alternative network specifications. The empirical analysis uses daily data for 16 sectorally diversified S$\&$P~500 stocks from 22~December~1998 to 20~October~2024.  
A one-step-ahead forecasting framework is implemented, and models are assessed using BIC, RMSFE, and MAFE, with forecasts evaluated against a single realised volatility proxy based on squared log-returns. The nine spatial weight matrices reflect diverse economic and statistical relationships, including Granger-filtered and EGARCH-based spillovers.  
Results show that some spatiotemporal models outperform standard GARCH benchmarks in out-of-sample forecasting accuracy. Notably, the Dynamic Spatiotemporal ARCH model achieves the lowest RMSFE and MAFE across all network specifications at minimal computational cost. Pairwise Diebold–Mariano tests confirm significant differences in predictive accuracy. These findings underscore the value of incorporating spatial structure into volatility modelling as a parsimonious and interpretable alternative for financial network analysis.

\end{abstract}


\onehalfspacing


\section{Introduction}

Volatility, broadly defined as the degree of variation in a process over time, is a central concept in financial econometrics. In financial markets, volatility reflects the magnitude and frequency of asset price fluctuations and plays a crucial role in asset pricing, risk management, derivative valuation, and investment strategies \cite{andersen2003modeling,hull2018options}.  
From a methodological perspective, volatility models can be classified into two main categories: models that specify the conditional variance of observed returns, such as the Generalised Autoregressive Conditional Heteroskedasticity (GARCH) family introduced by \cite{bollerslev1986generalized}, and latent-variable approaches like stochastic volatility \cite{taylor2008modelling,jacquier2002bayesian}.  
Over the past few decades, increasingly nonlinear extensions have been proposed to capture the complex behaviour of financial volatility, including multivariate formulations such as the Dynamic Conditional Correlation (DCC) model of \cite{engle2002dynamic} and the BEKK specification of \cite{engle1995multivariate}. 
These models have achieved widespread adoption due to their ability to capture time-varying volatility and co-movements across multiple assets. 
However, as the dimensionality of financial systems grows, so do the computational and interpretability challenges posed by multivariate volatility models. In response, recent research has explored the use of simplified structures that draw from spatial statistics, modelling inter-asset relationships via pre-defined spatial weight matrices or network adjacency structures. 

A prominent example is the proximity-based BEKK model of \cite{caporin2015proximity}, which imposes proportionality restrictions on the coefficient matrices using an economically meaningful proximity matrix. This yields a parsimonious formulation that reduces the dimensionality of the parameter space while preserving the conditional covariance matrix's positive definiteness. Although this specification embeds cross-sectional dependence indirectly—via the proximity matrix in the GARCH recursion, it does not allow for direct contemporaneous spatial spillovers in volatility. In contrast, spatiotemporal ARCH-type models explicitly introduce spatial lag terms, in which volatility at one location depends on the current or past volatility at neighbouring locations. Recent contributions in this direction include the spatiotemporal log-ARCH model \cite{sato2017spatial}, the dynamic spatiotemporal log-ARCH model \cite{otto2024dynamic}, the circular spatiotemporal GARCH model \cite{holleland2020stationary}, the spatial GARCH model of \cite{borovkova2012spatial}, and the newly proposed spatiotemporal EGARCH model \cite{chrisko2025exponential}. These frameworks allow volatility shocks to propagate over time and space, providing flexible yet computationally efficient alternatives to high-dimensional multivariate GARCH models. Their structural parsimony and modularity make them particularly well-suited for large financial networks and other high-dimensional time series applications. The key question remains: can these spatiotemporal approaches match—or surpass—the forecasting performance of traditional multivariate volatility models such as DCC and BEKK?
A good volatility model, as emphasized by \cite{engle2007good}, should exhibit several essential characteristics: it must accurately forecast volatility, capture mean-reversion and persistence (commonly referred to as volatility clustering, where large changes tend to be followed by large changes and small changes by small changes), and account for asymmetries in the impact of positive and negative shocks—known as the leverage effect. Additionally, it should allow for the inclusion of exogenous variables, be robust to leptokurtic (fat-tailed) return distributions, and effectively handle extreme market events. In modern applications, especially those involving high-frequency or high-dimensional data, computational scalability has become a critical requirement. Volatility models must not only be statistically sound but also efficient in terms of estimation and forecasting speed. 
When extended to spatiotemporal settings, models are further expected to capture contemporaneous spillovers across assets or spatial interactions. This added dimension of complexity enables a more realistic and informative analysis of interconnected systems. It is worth mentioning that, beyond finance, spatiotemporal volatility models have found applications across a variety of disciplines. For instance, \cite{holleland2020stationary} applies a spatiotemporal GARCH model to sea surface temperature anomalies in the Pacific Ocean to identify persistent climate patterns.

In this study, we evaluate a spectrum of volatility models ranging from fully dynamic and flexible frameworks—such as the Dynamic Conditional Correlation (DCC) and BEKK models to structurally simplified approaches inspired by spatial econometrics. The latter include spatiotemporal GARCH-type models that incorporate spatial lag terms on volatility, circular GARCH models, and proximity-based specifications. While DCC and BEKK dynamically estimate pairwise co-movements among assets in a data-driven manner, spatial volatility models rely on predefined or data-informed adjacency structures, offering potential advantages in terms of interpretability and computational efficiency.
The evaluation is made using the S$\&$P~500 index from 22 December 1998 to 20 October 2024, one of the most widely tracked assets in global financial markets. 
We assess model performance under various pre-defined network specifications, including correlation-based, economic distance, Piccolo-based, and volatility spillover matrices. We evaluate both in-sample model fit (via the Bayesian Information Criterion) and out-of-sample forecasting accuracy using standard metrics such as the Root Mean Squared Forecast Error (RMSFE) and the Mean Absolute Forecast Error (MAFE).
While comparing non-nested models poses challenges for formal inference, forecasting accuracy reflects the practical conditions under which financial models are typically employed, where the primary objective is to generate reliable predictions. 
Our findings highlight whether spatially structured volatility models can rival or exceed the performance of traditional multivariate approaches while offering improved interpretability and scalability.

The remainder of the paper is organised as follows. Section~\ref{sec2} introduces the volatility models under consideration, and outlines the framework for evaluating one-step-ahead forecasts. Section~\ref{sec3} details the construction of the spatial weight matrices, where spatial dependence reflects economic or statistical linkages rather than purely geographic proximity. Section~\ref{sec4} presents the S$\&$P~500 dataset and the empirical results, comparing predictive performance and computational efficiency across models. 
The final section concludes with implications for model selection in financial econometrics and suggests potential directions for future research.

\section{Brief description of volatility models in spatiotemporal statistics} \label{sec2}

In this section, we present the primary spatiotemporal volatility models selected for comparison, each chosen for its distinct methodological contribution to modelling dynamic interdependencies across space and time. These models naturally fall within the class of multivariate volatility models, given their application in high-dimensional contexts such as financial markets or regional economic systems, and or environmental systems by naturally accommodating systems where spatial structure or geographic references play a central role.
Multivariate volatility modelling is crucial for capturing cross-sectional interactions, such as co-movements among assets or spatial units. However, as the dimensionality increases, so does the computational complexity. This presents practical challenges in both estimation and interpretation, motivating the need for parsimonious yet flexible modelling frameworks.
Before introducing the specific spatiotemporal models under study, we briefly review two foundational multivariate GARCH models: the BEKK model and the Dynamic Conditional Correlation (DCC) model. These models are widely used benchmarks in the literature and provide a basis for understanding the extensions introduced in spatiotemporal settings.
The BEKK model, introduced by \cite{engle1995multivariate}, was designed to capture the full variance-covariance dynamics in a multivariate GARCH framework while ensuring positive definiteness by construction. On the other hand, the DCC model, developed by \cite{engle2001theoretical}, focuses on modelling time-varying conditional correlations in a computationally tractable manner. While BEKK is often employed in value-at-risk (VaR) forecasting and risk monitoring, DCC is especially suited for studying correlation dynamics, such as portfolio diversification or contagion effects across markets.
In general, variances and covariances are essential for forecasting risk thresholds such as VaR under regulatory frameworks (e.g., Basel Accord) \cite{caporin2012we}, while correlation structures are more relevant for portfolio allocation and hedging strategies. Both BEKK and DCC belong to the GARCH family but differ in structure and application scope. 

Let $(\boldsymbol{\varepsilon}_{t})'_{t\geq 0}$ denote the $n \times 1$ vector of residuals from a mean model (e.g., a Vector Autoregressive model), $(\boldsymbol{\varepsilon}_{1t}, \cdots, \boldsymbol{\varepsilon}_{nt} )'$. In their simplest form, these multivariate GARCH models are defined below.

\subsection{BEKK Model} \label{bekk_sec}

The BEKK model of \citet{engle1995multivariate} provides a flexible parametrisation of multivariate
volatility. For the BEKK(1,1) specification, the conditional covariance matrix is given by
\begin{equation}
\boldsymbol{\Sigma}_t
= \mathbf{C}\mathbf{C}' 
+ \mathbf{A}\,\boldsymbol{\varepsilon}_{t-1}\boldsymbol{\varepsilon}_{t-1}'\mathbf{A}'
+ \mathbf{B}\,\boldsymbol{\Sigma}_{t-1}\mathbf{B}',
\label{eq:bekk_short}
\end{equation}
where \( \mathbf{C} \) is lower triangular to ensure positive definiteness,  
\( \mathbf{A} \) captures the impact of past shocks,  
and \( \mathbf{B} \) governs persistence in conditional covariances. This formulation ensures that \( \boldsymbol{\Sigma}_t \) remains positive definite, which is crucial for its interpretation as a valid covariance matrix.
Although highly flexible, the unrestricted BEKK model contains \(O(n^2)\) parameters in each of the
coefficient matrices, making estimation computationally demanding as the dimension increases.
Consequently, empirical applications often adopt restricted versions such as diagonal, scalar, or
asymmetric BEKK \cite{ding1993long, kroner1998modeling}.
Estimation proceeds by maximum likelihood under the assumption of conditionally Gaussian
innovations, where the recursion \eqref{eq:bekk_short} is evaluated at each iteration of the
optimiser. Because the likelihood surface is high-dimensional and non-linear, derivative-based
numerical methods are typically employed to ensure stable and efficient convergence.

\subsection{DCC Model} \label{dcc_sec}

The Dynamic Conditional Correlation (DCC) model extends the Constant Conditional Correlation (CCC) model of \cite{bollerslev1990modelling} by allowing the conditional correlation matrix to evolve over time. In the DCC framework, the residuals are expressed as
\begin{eqnarray}
\boldsymbol{\varepsilon}_{t} &=& \mathbf{H}_{t}^{1/2} \mathbf{z}_{t}, \label{maineq} \\
\mathbf{H}_{t} &=& \mathbf{D}_{t}^{1/2} \mathbf{R}_{t} \mathbf{D}_{t}^{1/2}, \quad t = 1, 2, \dots, T,
\end{eqnarray}
where \( \boldsymbol{\varepsilon}_{t} \in \mathbb{R}^n \) is the vector of observations at time \( t \) with $n$ is the number of assets, 
    \( \mathbf{z}_{t} \sim \mathcal{N}(0, \mathbf{I}) \) are standardized i.i.d. innovations,
     \( \mathbf{H}_{t} \) is the conditional covariance matrix,
     \( \mathbf{D}_{t} \) is a diagonal matrix of time-varying variances,
    \( \mathbf{R}_{t} \) is the time-varying conditional correlation matrix.
The diagonal matrix \( \mathbf{D}_t \) contains the conditional variances: $\mathbf{D}_t = \text{diag}\left(h_{t}^{1/2}(s_1), h_{t}^{1/2}(s_2), \dots, h_{t}^{1/2}(s_n)\right)$,
with each \( h_t(s_i) \) following a univariate GARCH(1,1) process:
\begin{equation}
h_t(s_i) = \omega(s_i) + \alpha_i \varepsilon_{t-1}^2(s_i) + \beta_i h_{t-1}(s_i),
\end{equation}
where \( \omega(s_i) > 0 \), \( \alpha_i, \beta_i \geq 0 \), and \( \alpha_i + \beta_i < 1 \) to ensure positivity and covariance stationarity.
The term $\omega(s_i)$ represents the baseline volatility of stock $i$, capturing its long-run average variance without external shocks. The second component, $\textbf{R}_{t}$, is the $n\times n$ conditional correlation matrix, capturing the dynamic dependence structure between the series. It is derived from the $n \times n$ quasi-correlation matrix $\textbf{Q}_{t}$, given by 
  \begin{align}
    \textbf{R}_{t} &= \text{diag}(\textbf{Q}_{t})^{-1/2} \textbf{Q}_{t} \text{diag}(\textbf{Q}_{t})^{-1/2}, \label{eq:correlation} \\
    \textbf{Q}_{t} &= (1 - \lambda_1 - \lambda_2)\overline{\mathbf{Q}} + \lambda_1 \boldsymbol{\tilde{u}}_{t-1}\boldsymbol{\tilde{u}}_{t-1}^\prime + \lambda_2 \textbf{Q}_{t-1}, \label{eq:quasicorrelation}
  \end{align}
where $\overline{\mathbf{Q}}$ is the long-run (unconditional) covariance matrix of the standardised residuals. These standardised residuals are given by $\tilde{\boldsymbol{u}}_t = \textbf{D}_t^{-1/2} \boldsymbol{u}_t = (\frac{\epsilon_{t}(s_1)}{h_{t}^{1/2}(s_1)}, \frac{\epsilon_{t}(s_2)}{h_{t}^{1/2}(s_2)}, \ldots, \frac{\epsilon_{t}(s_n)}{h_{t}^{1/2}(s_n)})
$, an $n \times 1$ vector. The parameters $\lambda_1$ and $\lambda_2$ determine correlations short-term and long-term behaviour, respectively.
The parameters of the DCC model are estimated using the two-stage method.  First, the univariate GARCH parameters for each series are estimated independently using Maximum Likelihood Estimation (MLE). The standardised residuals, $\tilde{u}_{t}(s_i)$, are then derived and used to estimate the correlation parameters ($\lambda_1, \lambda_2$) via MLE. For further details, see \cite{engle_theoretical_2001}.

The DCC and BEKK families have inspired numerous extensions aimed at enhancing flexibility, reducing computational complexity, and capturing empirical features such as asymmetry and cross-sectional spillovers. Scalar, Diagonal, and Full versions of both models differ in how they parameterise the conditional covariance matrix: scalar forms are the most parsimonious, while full BEKK models offer the most general structure at the cost of a substantial increase in the number of parameters. Hadamard-type variants introduce element-wise operations that balance flexibility with tractability \cite{caporin2012we}. Further developments include asymmetric formulations such as the Asymmetric DCC (ADCC) model \cite{cappiello2006asymmetric}, and the Generalised DCC (GDCC) model, which allows for richer correlation dynamics. GARCH-X models \cite{engle2008spline} incorporate exogenous covariates and can be embedded within both DCC and BEKK specifications. In DCC models, the reported number of parameters often omits the univariate GARCH components, typically at least $3k$ for $k$ assets and grows considerably when asymmetric or leverage effects are modelled \cite{aielli2013dynamic, engle2012dynamic}. A fairly detailed comparison of these models is provided in \cite{caporin2012we}.
We now shift focus to the spatiotemporal analogues of these models. The dynamic spatiotemporal ARCH model captures both local temporal dynamics and spatial dependencies, offering a flexible approach to modelling volatility clustering. The proximity-structured multivariate volatility model integrates spatial weighting directly into the covariance structure, enabling the explicit modelling of spatial spillovers. The stationary spatiotemporal GARCH model extends classical GARCH formulations to account for joint spatial-temporal stationarity, making it well-suited for long-horizon forecasting. Finally, the spatiotemporal EGARCH model introduces asymmetric effects and allows for instantaneous volatility transmission across networks, capturing the interplay between spatial proximity and shock directionality under different regimes.

All models considered in this study follow the same multiplicative structure in (\ref{maineq}), where the matrix $\mathbf{H}_t$ represents the conditional volatility structure, encompassing both variances and covariances. 
In the spatiotemporal context, $\mathbf{H}_t$ is often simplified to a diagonal matrix $\mathrm{diag}(\mathbf{h}_t^{1/2})$, with 
$\mathbf{h}_t^{1/2} = (h_t^{1/2}(s_1), \dots, h_t^{1/2}(s_n))^\top$ 
containing the conditional standard deviations at each location~$s_i$. 
Alternatively, $\mathbf{H}_t$ may denote a full conditional covariance matrix~$\boldsymbol{\Sigma}_t$ as in~(\ref{eq:bekk_short}). 
The innovation vector $\boldsymbol{z}_t = (z_t(s_1), \dots, z_t(s_n))^\top$ consists of i.i.d. random variables with zero mean and unit variance across both dimensions. 
Note that volatility typically refers to the square root of the conditional variance. 
In some model formulations, particularly those expressed in logarithmic or reduced forms $\mathbf{H}_t$ serves as a parametric representation of the conditional variance–covariance structure rather than the exact conditional covariance matrix 
$\mathrm{Var}[\boldsymbol{\varepsilon}_t \mid \mathcal{F}_{t-1}]$. 
In what follows, we focus on basic spatiotemporal volatility models, excluding exogenous covariates for clarity, and note that \( \xmat{W} \) is an \( n\times n\) spatial weight matrix.

\subsection{Proximity-structured multivariate volatility model (STBEKK)} \label{stbekk_sec}

The first model is that of \cite{caporin2015proximity}, who propose a structured or restricted volatility model to address the dimensionality curse in multivariate volatility modelling by incorporating a structured parametrisation based on economic proximity. The approach uses weight matrices to model conditional covariances as with classic spatial models, reducing the complexity of parameter estimation while preserving the interpretability of economic relationships. Their framework is particularly suited for financial networks as it integrates information about economic or spatial connections to enhance the analysis of multivariate volatility dynamics. The model equation is given as follows 
\begin{equation}
\boldsymbol{\Sigma}_t = \xmat{C} + \xmat{A} \xvec{\varepsilon}_{t-1} \xvec{\varepsilon}_{t-1}' \xmat{A}' + \xmat{B} \boldsymbol{\Sigma}_{t-1} \xmat{B}',
\end{equation}
where \(\boldsymbol{\Sigma}_t\) is the matrix \(n \times n\) of covariances, \(\xmat{A}\), \(\xmat{B}\), and \(\xmat{C}\) are matrices of unrestricted parameters \(n \times n\). To reduce the number of parameters from \( \mathcal{O}(n^4) \) to \( \mathcal{O}(n) \), \cite{caporin2015proximity} introduce a proximity-based parameterisation
\begin{eqnarray}
\xmat{C} &=& \xmat{S}^{-1} V \xmat{S}^{-1'}, \quad \xmat{S} = \xmat{I}_n - \xmat{S}_1 \xmat{W}, \quad \xmat{S}_1 = \operatorname{diag}(s^{(1)}), \quad \xmat{V} = \operatorname{diag}(v),    \\
\xmat{A} &=& \xmat{A}_0 + \xmat{A}_1 \xmat{W}, \quad \xmat{A}_j = \operatorname{diag}(\alpha^{(j)}), \quad \xmat{B} = \xmat{B}_0 + \xmat{B}_1 W, \quad \xmat{B}_j = \operatorname{diag}(\beta^{(j)}),
\end{eqnarray}
for \( j = 0, 1 \), where \( s^{(1)}, v, \alpha^{(j)}, \) and \( \beta^{(j)} \) are \(n \times 1\) vectors of unknown parameters.
This structured specification allows for both own and cross-unit effects in the conditional covariance dynamics through the inclusion of \( \xmat{W} \), which encodes proximity relationships among the entities. A homogeneous parameter version can be obtained by assuming parameter vectors are constant across units. In this case, the total number of parameters reduces dramatically; in the group-homogeneous specification presented by \cite{caporin2015proximity}, only six parameters are required.
Estimation is carried out via quasi-maximum likelihood, and asymptotic properties of the estimator are examined numerically. The spillover structure arising in the terms \( \mathbf{A} \boldsymbol{\varepsilon}_{t-1} \boldsymbol{\varepsilon}_{t-1}' \mathbf{A}' \) and \( \mathbf{B} \boldsymbol{\Sigma}_{t-1} \mathbf{B}' \) is directly governed by the choice of the proximity matrix \( \mathbf{W} \), which allows the model to flexibly adapt to different network or spatial configurations.

Although not originally formulated as a spatiotemporal model, the proximity-structured volatility model shares key features with spatiotemporal frameworks by encoding similarity across units through a proximity matrix. This matrix can represent geographic, economic, or statistical relationships, making the model particularly flexible for applications involving latent spatial or network structures. As such, it serves as a conceptual bridge between classical multivariate GARCH models and more recent spatial or network-based volatility models.
In their implementation, \cite{caporin2015proximity} recommends experimenting with different types of proximity matrices. One approach that is particularly appealing in financial applications is to define weights based on sectoral affiliation, that is, whether two stocks belong to the same economic sector. This allows the spatial weight matrix to reflect economically meaningful groupings through a network structure derived from group membership.

\subsection{Dynamic Spatiotemporal ARCH model (DSTARCH)} \label{dstarch_sec}
The first attempts to introduce spatial dependence into ARCH-type volatility models were made by \citet{otto2018generalised} and independently by \citet{sato2021spatial}, who proposed spatial log-ARCH formulations in which the logarithmic conditional variance at a given location depends instantaneously on the variances of neighbouring locations. These models marked a shift from traditional multivariate GARCH approaches by embedding spatial interactions directly into the volatility mechanism rather than modelling cross-asset dependence through full covariance matrices.
To incorporate temporal dynamics alongside spatial effects, \citet{otto2024dynamic} introduced the Dynamic Spatiotemporal ARCH (DST-ARCH) model, in which the log-volatility is a function of the lagged log-squared outcome variable, the spatial lag of the log-squared outcome variable, and a spatiotemporal interaction term.  In its simplest form, without exogenous regressors, the model for the log-volatility vector 
$\boldsymbol{H}^{*}_t = (\log h_t(s_1), \ldots, \log h_t(s_n))'$ is given by
\begin{equation} \label{eq:networkArch}
    \boldsymbol{H}^{*}_t = \boldsymbol{\omega} + \boldsymbol{\Gamma} \boldsymbol{\varepsilon}^{*}_{t-1} + \rho \mathbf{W} \boldsymbol{\varepsilon}^{*}_t,
\end{equation}
where \( \rho \in \mathbb{R} \) captures the strength of contemporaneous (instantaneous) volatility spillovers across the network. The vector \( \boldsymbol{\omega} = (\omega_1, \omega_2, \dots, \omega_n)' \) contains asset-specific baseline log-volatilities, while \( \boldsymbol{\Gamma} = \mathrm{diag}(\gamma_1, \gamma_2, \dots, \gamma_n) \) is a diagonal matrix capturing stock-specific temporal ARCH effects. $\boldsymbol{\varepsilon}^{*}_t$ in this case represents the pairwise logarithm of the vector of observations $(\log \varepsilon_t(s_1), \ldots, \log \varepsilon_t(s_n))'$.
The total number of free parameters in this simplified version is \( 2n + 1 \), accounting for \( n \) intercepts, \( n \) temporal coefficients, and the single network parameter \( \rho \). A statistically significant \( |\rho| \) implies that shocks propagate instantaneously through the network, rather than exclusively via temporal lags. This is a key distinguishing feature compared to network GARCH-type models, which allow only lagged interactions.  While effective in modelling spatiotemporal volatility clustering, a notable limitation of the DST-ARCH model is its reliance on instantaneous spatial dependence, which constrains its forecasting capacity. 
Following \cite{mattera2024network}, the model admits an ARMA-type representation that facilitates estimation using orthonormal transformations and the Generalised Method of Moments (GMM). The one-step-ahead forecast of log-volatility is then given by
\begin{equation} \label{eq:networkForecast}
    \boldsymbol{H}^{*}_{t+1} = (\mathbf{I}_n - \hat{\rho} \mathbf{W})^{-1} \left[ \hat{\boldsymbol{\Gamma}} \boldsymbol{\varepsilon}^{*}_{t} +  \hat{\boldsymbol{\phi}}_0 \right],
\end{equation}
where \( \hat{\boldsymbol{\phi}}_0 \) denotes the estimated fixed effect from the orthonormal transformation, and \( \hat{\rho} \), \( \hat{\boldsymbol{\Gamma}} \) are consistent GMM estimates. Specifically, the transformation matrix $\mathbf{Q}$ satisfies $\mathbf{QWQ}' = \boldsymbol{\Lambda}$, where $\boldsymbol{\Lambda}$ is diagonal. This reduces the $n$-dimensional dependence induced by $\mathbf{W}$ to $n$ independent scalar equations, greatly simplifying estimation.

\subsection{Circular spatiotemporal GARCH model (STGARCH)} \label{stgarch_sec}
The spatiotemporal GARCH model introduced by \cite{holleland2020stationary} extends the classical GARCH framework by incorporating both spatial and temporal dependencies. It is defined on the domain $\mathbb{Z} \times \mathbb{Z}^d$, where $d \geq 1$ denotes the spatial dimension. The model captures interactions over space and time by introducing a structured dependence on past values and volatilities of neighbouring spatial locations. To address boundary issues inherent in finite spatial domains, the authors also propose a circular version of the model, well-suited for spatially periodic structures or replicated spatial patterns.

The general form of the model is given by:
\begin{equation}
h_t^2(\mathbf{s}) = \omega + \sum_{i=1}^p \sum_{\mathbf{v} \in \Delta_{1i}} \alpha_i(\mathbf{v}) Y_{t-i}^2(\mathbf{s} -\mathbf{v}) + \sum_{i=1}^q \sum_{\mathbf{v} \in \Delta_{2i}} \beta_i(\mathbf{v}) h_{t-i}^2(\mathbf{s}-\mathbf{v}),
\end{equation}
where $\omega > 0$ is a scalar parameter, and $\alpha_i$, $\beta_i$ are parameter functions with finite support defined over $\mathbb{Z}^d$. The support sets are defined as:
\begin{equation}
\Delta_{1i} = \{\mathbf{v} \in \mathbb{Z}^d : \alpha_i(\mathbf{v}) > 0\}, \quad 
\Delta_{2i} = \{\mathbf{v} \in \mathbb{Z}^d : \beta_i(\mathbf{v}) > 0\},
\end{equation}
and the model order $(p,q)$ corresponds to the maximum lag for which the respective supports are non-empty. The parameter set is denoted $\Theta = (\omega, \boldsymbol{\alpha}, \boldsymbol{\beta})$ with $\boldsymbol{\alpha} = \{\alpha_i(\mathbf{v})\}$ and $\boldsymbol{\beta} = \{\beta_i(\mathbf{v})\}$ collected over all relevant lags and spatial shifts.
The circular version replaces $\mathbb{Z}^d$ by the quotient group $\mathcal{R} = \mathbb{Z}^d / (m\mathbb{Z}^d)$, for some $m \in \mathbb{Z}_+^d$, effectively wrapping the spatial domain onto a torus. This makes $\{Y_t(\mathbf{s})\}$ a process indexed over $\mathbb{Z} \times \mathcal{R}$, where spatial operations like shifts and additions are performed modulo $m$, ensuring translation-invariant neighbourhoods and preserving stationarity. \cite{holleland2020stationary} derive strong consistency and asymptotic normality of the quasi-maximum likelihood estimator (QMLE) for the model parameters. The circular formulation offers a theoretically elegant and computationally feasible approach to modelling spatiotemporal volatility, especially in gridded data contexts like environmental or climatological studies.

\subsection{Spatial GARCH expressed as spGARCH-X} \label{garch_X_sec}

The Spatial GARCH(1,1) model extends the traditional GARCH framework by incorporating network-based volatility spillovers through a spatial weight matrix, making it suitable for multivariate conditional variance modelling. In this setup, each asset’s conditional variance depends not only on its own past squared returns and past variances but also on the spatially weighted contributions from neighbouring assets. Specifically, lagged squared returns and lagged conditional variances of the $N-1$ other stock markets are aggregated using a spatial weight matrix, resulting in two spatial exogenous variables and two additional parameters, $a_{2, i}$ and $b_{2, i}$, which quantify the influence of these spatial aggregates on asset $i$'s volatility.
Let $\mathbf{W}$ be a fixed, non-stochastic spatial weight matrix of dimension $N \times N$, encoding the strength of volatility transmission between assets $s_i$ and $s_j$. Assuming the initial innovations vector $\boldsymbol{\varepsilon}_0$ is observable, the Spatial GARCH(1,1) model proposed by \cite{borovkova2012spatial} is specified as 
\begin{equation} \label{Eq:borov}
\mathbf{H}_t = \mathbf{A}_0 + (\mathbf{A}_1 + \mathbf{A}_2 \mathbf{W}) \boldsymbol{\varepsilon}_{t-1}^2 + (\mathbf{B}_1 + \mathbf{B}_2 \mathbf{W}) \mathbf{H}_{t-1},
\end{equation}
where $\boldsymbol{\varepsilon}_t^2 = (\varepsilon_{t,1}^2, \ldots, \varepsilon_{t,N}^2)^\top$ denotes the vector of squared residuals, $\mathbf{A}_0 = (a_{0,1}, \ldots, a_{0,N})^\top$ is a vector of intercept terms, and for $k = 1,2$, $\mathbf{A}_k = \text{diag}(a_{k,1}, \ldots, a_{k,N})$ and $\mathbf{B}_k = \text{diag}(b_{k,1}, \ldots, b_{k,N})$ are diagonal matrices of non-negative parameters. All coefficients and spatial weights are constrained to be non-negative to ensure the positivity of the conditional variances.
The parameters $a_{1,i}$ and $b_{1,i}$ retain their classical GARCH(1,1) interpretation, where $a_{1,i}$ captures sensitivity to recent shocks and $b_{1,i}$ reflects the persistence of volatility. The spatial terms $a_{2, i}$ and $b_{2, i}$ extend the model by capturing short- and long-term spillovers from neighbouring assets. A large $a_{2, i}$ implies that asset $i$ is more reactive to recent innovations in neighbouring nodes, while a large $b_{2, i}$ indicates that persistent volatility in neighbouring nodes influences asset $i$. 

The model parameters are estimated using a conditional quasi-maximum likelihood (QMLE) procedure as proposed in \cite{borovkova2012spatial}. Assuming that the residuals $z_t(s_i)$ are conditionally independent across locations and time, the joint likelihood decomposes into a product of location-specific likelihoods. Each asset is treated as a univariate GARCH(1,1) model augmented with two spatially lagged exogenous variables given by 
\begin{eqnarray} \label{eq:spatial_garchx}
h_{i,t} &=& a_{0,i} + a_{1,i} \varepsilon_{i,t-1}^2 + b_{1,i} h_{i,t-1} + a_{2,i} X_{i,t-1} + b_{2,i} Y_{i,t-1},
\end{eqnarray}
where the spatial lag terms are defined as
\begin{eqnarray}
X_{i,t-1} &=& \sum_{j=1}^{N} w_{ij} \varepsilon_{j,t-1}^2, \qquad Y_{i,t-1} = \sum_{j=1}^{N} w_{ij} h_{j,t-1},
\end{eqnarray}
The model is called GARCH-X(1,1) because it includes two exogenous terms: \( X_{i,t-1} \), which aggregates past squared returns from neighbouring units, and \( Y_{i,t-1} \), which aggregates their lagged variances. The estimation is carried out iteratively. Initially, univariate GARCH(1,1) models are fitted to each asset to obtain starting values for the parameters and conditional variances. These estimates are then updated iteratively: at each step, spatially lagged exogenous variables are recalculated based on the previous iteration’s estimates, and each asset’s parameters are re-estimated using MLE. This process continues until convergence. Each unit has five parameters; hence, for \( N \) units, the total number of parameters to estimate is \( 5N \).

\subsection{Spatiotemporal EGARCH Model (STEGARCH)} \label{stegarch_sec}

The Spatiotemporal EGARCH model of \cite{chrisko2025exponential} extends the classical EGARCH \cite{Nelson91} framework by incorporating spatial interactions and asymmetric shock effects within a log-volatility representation. Let $\mathbf{H}_t^* = (\log h_t(s_1), \ldots, \log h_t(s_n))'$ denote the vector of log-conditional variances. The model is specified as
\begin{equation} \label{eq:spegarch}
\mathbf{H}_t^*
= \boldsymbol{\alpha}_1
+ \rho_0\, \mathbf{W}_1 g(\boldsymbol{\varepsilon}_t)
+ \rho_1\, g(\boldsymbol{\varepsilon}_{t-1})
+ \lambda_0\, \mathbf{W}_2 \mathbf{H}_t^*
+ \lambda_1\, \mathbf{H}_{t-1}^*,
\end{equation}
where $g(\boldsymbol{\varepsilon}_t)$ is the EGARCH-type nonlinear transformation given by 
\begin{equation}
g(\boldsymbol{\varepsilon}_t)
= \Theta\, \boldsymbol{\varepsilon}_t
+ \xi\left(|\boldsymbol{\varepsilon}_t| - \mathbb{E}|\boldsymbol{\varepsilon}_t|\right),
\end{equation}
which captures both signed and magnitude-driven asymmetries.  
Because the EGARCH transformation contains a signed term ($\Theta \boldsymbol{\varepsilon}_t$) together with an asymmetric magnitude component, contemporaneous spillovers transmitted through $\mathbf{W}_1$ may themselves be asymmetric. Economically, this allows shocks at one asset to have instantaneous and directionally different effects on the volatility of neighbouring assets.
The parameters $\rho_0$ and $\rho_1$ govern spatial and temporal asymmetric effects, while $\lambda_0$ and $\lambda_1$ describe spatial and temporal persistence in log-volatility. The matrices $\mathbf{W}_1$ and $\mathbf{W}_2$ are exogenous $n\times n$ spatial weight matrices, typically reflecting economically meaningful relations such as sectoral proximity, cross-market linkages, or correlation-based neighbourhoods.
Estimation proceeds by quasi-maximum likelihood. Because $\mathbf{H}_t^*$ appears on both sides of \eqref{eq:spegarch}, every likelihood evaluation requires solving the linear system
\begin{equation}
(\mathbf{I}_n - \lambda_0 \mathbf{W}_2)\mathbf{H}_t^*
= \boldsymbol{\alpha}_1
+ \rho_0\, \mathbf{W}_1 g(\boldsymbol{\varepsilon}_t)
+ \rho_1\, g(\boldsymbol{\varepsilon}_{t-1})
+ \lambda_1\, \mathbf{H}_{t-1}^*,
\end{equation}
so, numerical inversion is needed at each iteration. Instability may arise when $\lambda_0$ approaches the boundary of invertibility, as reflected in the model's empirical performance.
Despite these numerical challenges, the STEGARCH specification offers a compact yet flexible framework for capturing contemporaneous and lagged spatial volatility spillovers, while explicitly including asymmetric shock transmission across the network.

A wide range of spatiotemporal volatility models has been proposed in the literature to capture both temporal and cross-sectional dependencies in financial time series. These models differ in how they incorporate spatial interactions—whether through network-based feedback, exogenous spatial regressors, or structured multivariate frameworks. Table~\ref{tab:volatility_models} provides a comparative overview of key models, including the DSTARCH, various spatiotemporal GARCH extensions, the spatiotemporal EGARCH capturing asymmetries, as well as spatial GARCH-X models that incorporate lagged variances and returns from neighbouring units. Also included are benchmark multivariate specifications such as the DCC and BEKK models, and the spatially structured BEKK by \cite{holleland2020stationary}, which is notable for its stationarity on spatial grids with circular boundary conditions. Together, these models offer a flexible toolkit for analysing volatility dynamics in interconnected financial systems.

\begin{table}[h!]
\small
\centering
\caption{Conditional volatility specifications for the models considered. 
“Cont.~dep.” indicates contemporaneous network dependence via the matrix $\xmat{W}$.}
\label{tab:volatility_models}
\begin{tabularx}{\textwidth}{lXcc}
\toprule
\textbf{Model} & \textbf{Specification } & \textbf{Cont.~dep. } & \textbf{Volatility form} \\
\midrule
\textbf{DCC} & 
$\mathbf{H}_t = \mathbf{D}_t\,\mathbf{R}_t\,\mathbf{D}_t$,  
with $\mathbf{R}_t$ evolving via a dynamic correlation process.  
Captures time-varying conditional correlations in a two-step framework.\ref{dcc_sec}  & No & Covariance \\[.5cm]

\textbf{BEKK}  & 
$\mathbf{H}_t = \mathbf{C}'\mathbf{C} + \mathbf{A}'\boldsymbol{\varepsilon}_{t-1}\boldsymbol{\varepsilon}_{t-1}'\mathbf{A} + \mathbf{B}'\mathbf{H}_{t-1}\mathbf{B}$.  
Ensures positive-definite covariance matrices through a flexible parametrisation.\ref{bekk_sec} & No & Covariance \\[.5cm]

\textbf{Proximity-structured BEKK} & 
$\mathbf{H}_t = \mathbf{C}'\mathbf{C} + A(\mathbf{W})'\boldsymbol{\varepsilon}_{t-1}\boldsymbol{\varepsilon}_{t-1}'A(\mathbf{W}) + B(\mathbf{W})'\mathbf{H}_{t-1}B(\mathbf{W})$,  
where $A(\mathbf{W})=\alpha\mathbf{I}+\beta\mathbf{W}$ and $B(\mathbf{W})=\gamma\mathbf{I}+\delta\mathbf{W}$.  
Imposes structure via a proximity matrix to reduce dimensionality.\ref{stbekk_sec} & No & Covariance \\[.5cm]

\textbf{Dynamic Network Log-ARCH}  & 
$\mathbf{H}^{*}_t = \boldsymbol{\omega} + \boldsymbol{\Gamma}\boldsymbol{\varepsilon}^{*}_{t-1} + \rho\,\mathbf{W}\boldsymbol{\varepsilon}^{*}_t$.  
Log-volatility model with lagged and contemporaneous network effects, capturing instantaneous volatility spillovers.\ref{dstarch_sec} & Yes & Log-variance \\[.5cm]

\textbf{STGARCH } & 
$\xmat{H}_t = \boldsymbol{\omega} + 
\mathbf{A}(\xmat{W})\,\xvec{\varepsilon}_{t-1}^2 + 
\mathbf{B}(\mathbf{W})\,\mathbf{H}_{t-1}$.  
Spatiotemporal extension allowing both temporal and spatial dependence.\ref{stgarch_sec} & No & Variance \\[.5cm]

\textbf{Spatial GARCH-X}  & 
$\mathbf{H}_t = \xmat{A}_0 + (\xmat{A}_1+\xmat{A}_2\xmat{W})\xvec{\varepsilon}_{t-1}^2 + (\xmat{B}_1+\xmat{B}_2\xmat{W})\xmat{H}_{t-1} $.  
Extends GARCH by adding spatial lags of past shocks and volatilities.\ref{garch_X_sec} & No & Variance   \\[.5cm]

\textbf{Spatiotemporal EGARCH}  & 
$\xmat{H}^{*}_t = \boldsymbol{\alpha} + 
\rho_0\,\xmat{W}g(\boldsymbol{\varepsilon}_t) + 
\rho_1\,g(\boldsymbol{\varepsilon}_{t-1}) +
\lambda_0\,\mathbf{W}\xmat{H}^{*}_t + 
\lambda_1\,\xmat{H}^{*}_{t-1}$.  
Models asymmetric spillovers and allows contemporaneous spatial dependency.\ref{stegarch_sec} & Yes & Log-variance  \\

\bottomrule
\end{tabularx}
\end{table}

Finally, because our models explicitly incorporate spatial dependence, the construction of spatial weight matrices is central to both estimation and forecasting. These matrices define the nature and intensity of spatial spillovers and will be discussed in detail in the next section.

\section{Construction of weight matrices for financial networks}\label{sec3}

Financial networks provide a flexible and informative representation of cross-sectional dependence among assets. Nodes correspond to individual stocks, while edges encode economic relationships such as comovement, predictive influence, or volatility spillovers \cite{baumohl2018networks}. To incorporate these interactions into spatiotemporal volatility models, we construct a family of spatial weight matrices $\xmat{W}$ that capture different notions of similarity or directional influence between stocks. Unlike classical spatial econometrics, where proximity is tied to geographic distance, financial networks depend on behavioural similarity derived from return dynamics.

We construct ten matrices in total, grouped into four classes reflecting increasing levels of modelling complexity:  
(i) fully connected similarity-based networks,  
(ii) Granger-filtered directed networks,  
(iii) sparse $k$-nearest neighbour networks, and  
(iv) a spillover-based matrix derived from EGARCH dynamics.  
These matrices allow us to evaluate how alternative representations of cross-sectional structure affect the performance of spatiotemporal volatility models.

\paragraph{Similarity-based fully connected matrices.}
Following \citet{mattera2024network}, we first construct three symmetric, fully connected matrices based on alternative distance measures between stock residuals. Let $\varepsilon_t(s_i)$ denote the residuals from a mean model for stock $i$. We consider the Euclidean distance, the correlation-based distance, and the Piccolo AR distance \cite{piccolo_distance_1990}:
\[
d_{ij}^{(E)} = \sqrt{\sum_{t=1}^{T} (\varepsilon_t(s_i)-\varepsilon_t(s_j))^2}, \qquad
d_{ij}^{(C)} = \sqrt{2(1-\rho_{ij})}, \qquad
d_{ij}^{(P)} = \sqrt{\sum_{p=1}^P(\gamma_{ip}-\gamma_{jp})^2}.
\]
Here $\rho_{ij}$ is the Pearson correlation between returns of $i$ and $j$, while $\gamma_{ip}$ denotes the autoregressive coefficients of order $p$ fitted to the log-squared residuals of stock $i$. In the Piccolo metric, lag order $P$ is truncated using an information criterion.\footnote{If two series have different optimal lag orders $P_i$ and $P_j$, we set $P = \max(P_i, P_j)$ and pad with zeros as in \cite{piccolo_distance_1990}.}

Weights are defined as inverse distances,
\[
w_{ij} = \begin{cases}
1 / d_{ij}, & i \neq j,\\
0, & i=j,
\end{cases}
\]
and row-normalized to ensure $\sum_j w_{ij} = 1$. This yields the matrices
$\mathbf{W}^{\mathrm{E}},\ \mathbf{W}^{\mathrm{C}},\ \mathbf{W}^{\mathrm{P}}$.

\paragraph{Granger-filtered directed matrices.}
The fully connected matrices above describe similarity but not predictability. To introduce directionality and remove spurious connections, we apply pairwise Granger causality tests \cite{granger_investigating_1969,papana2017financial}. For each ordered pair $(i,j)$, we test whether $j$ Granger-causes $i$ using residual-based VAR models. A binary filter is defined as
\[
g_{ij} =
\begin{cases}
1, & \text{if } j \text{ Granger-causes } i \ (\mathrm{p}<0.05),\\
0, & \text{otherwise}.
\end{cases}
\]
Filtered weights are obtained via element-wise multiplication,
\[
\tilde{w}_{ij} = w_{ij} g_{ij}, \qquad \tilde{w}_{ii}=0,
\]
followed by row normalisation. This produces the directed matrices
$\mathbf{W}^{\mathrm{E\text{-}G}},\ \mathbf{W}^{\mathrm{C\text{-}G}},\ \mathbf{W}^{\mathrm{P\text{-}G}}$.

\paragraph{Sparse 5-nearest neighbour matrices.}
To impose sparsity and mitigate the influence of distant nodes, we construct $k$-nearest neighbour (k-NN) graphs with $k=5$ for each distance definition. For each asset $i$, we identify the five closest neighbours under $d_{ij}^{(E)}$, $d_{ij}^{(C)}$, and $d_{ij}^{(P)}$. We set
\[
w_{ij}^{(k\text{-NN})} =
\begin{cases}
1/5, & j \text{ is among the 5 nearest neighbours of } i,\\
0, & \text{otherwise}.
\end{cases}
\]
These matrices are symmetric and already row-stochastic, yielding
$\mathbf{W}^{\mathrm{E\text{-}NN}},\ \mathbf{W}^{\mathrm{C\text{-}NN}},\ \mathbf{W}^{\mathrm{P\text{-}NN}}$.

\paragraph{EGARCH-based volatility spillover matrix.}
To construct a network explicitly based on volatility spillovers, we estimate univariate EGARCH(1,1) models for each asset,
\[
\log h_{i,t} = \omega_i + \beta_i \log h_{i,t-1} + \alpha_i |z_{i,t-1}| + \gamma_i z_{i,t-1},
\qquad
z_{i,t} = \varepsilon_{i,t} / \sqrt{h_{i,t}},
\]
and extract the conditional volatilities $\sigma_{i,t}=\sqrt{h_{i,t}}$. Granger tests using VAR models on the volatility series determine the direction and intensity of spillovers. We define
\[
w_{ij}^{\mathrm{Spill}} =
\begin{cases}
1 - p_{ij}, & j \text{ Granger-causes } i,\\
0, & \text{otherwise},
\end{cases}
\]
where $p_{ij}$ is the Granger-test p-value. The resulting matrix $\mathbf{W}^{\mathrm{Spill}}$ is asymmetric and measures the strength of volatility transmission.

\medskip
Together, these matrices allow us to examine spatiotemporal volatility models under diverse network topologies. The progression from simple similarity measures to directed spillover structures reflects increasing modelling complexity and financial interpretability. A complete specification is provided in Appendix~\ref{fig:network_grid}, and a summary appears in Table~\ref{sum_weights}.

\begin{table}[htbp]
\centering
\caption{Overview of constructed weight matrices} \label{sum_weights}
\begin{tabular}{@{}lllll@{}}
\toprule
Matrix ID & Description & Symmetric & Sparse & Directional \\ \midrule
$\mathbf{W}^{\text{Euclidean}}$ & Inverse Euclidean distance     & \ding{51} & \ding{55} & \ding{55} \\
$\mathbf{W}^{\text{Corr}}$      & Inverse correlation distance   & \ding{51} & \ding{55} & \ding{55} \\
$\mathbf{W}^{\text{Piccolo}}$   & Inverse AR (Piccolo) distance  & \ding{51} & \ding{55} & \ding{55} \\
$\mathbf{W}^{\text{E-G}}$       & Euclidean + Granger filter     & \ding{55} & \ding{55} & \ding{51} \\
$\mathbf{W}^{\text{C-G}}$       & Correlation + Granger filter   & \ding{55} & \ding{55} & \ding{51} \\
$\mathbf{W}^{\text{P-G}}$       & Piccolo + Granger filter       & \ding{55} & \ding{55} & \ding{51} \\
$\mathbf{W}^{\text{E-NN}}$      & 5-NN based on Euclidean        & \ding{51} & \ding{51} & \ding{55} \\
$\mathbf{W}^{\text{C-NN}}$      & 5-NN based on correlation      & \ding{51} & \ding{51} & \ding{55} \\
$\mathbf{W}^{\text{P-NN}}$      & 5-NN based on Piccolo          & \ding{51} & \ding{51} & \ding{55} \\
$\mathbf{W}^{\text{Spill}}$     & EGARCH Granger spillovers      & \ding{55} & \ding{55} & \ding{51} \\ \bottomrule
\end{tabular}
\end{table}

\section{Empirical analysis and results: S$\&$P  500 Dataset} \label{sec4} 
This section presents the empirical analysis based on daily returns for sixteen S$\&$P~500 stocks. We begin by outlining the forecasting evaluation framework used to assess predictive accuracy across different spatial specifications. We then describe the dataset and report empirical results across the various weight matrices, highlighting how spatial structure influences volatility forecasting in a financial network setting.
 
\subsection{Forecasting Evaluation}

Evaluating the performance of volatility models poses a fundamental challenge because true volatility is latent and not directly observable. A common workaround is to use historical volatility, typically computed as the standard deviation of returns over a rolling window. In simplified settings—such as when the conditional mean is assumed to be zero—squared returns are often used as a proxy. More sophisticated methods leverage high-frequency data to construct realised volatility measures, as proposed by \citet{andersen2001distribution} and \citet{barndorff2001non}; see also \citet{eckel2011measuring} for a broader discussion. However, such methods are not employed here due to data limitations.
Another key issue in volatility forecast evaluation is the choice of loss function. As noted by \citet{bollerslev1992arch} and \citet{diebold19968}, the selection of an appropriate loss function depends heavily on the modelling objective. \citet{hansen2005forecast} compared six loss functions, demonstrating that model rankings can vary substantially depending on the chosen metric. In this study, we focus on two commonly used criteria: the Root Mean Squared Forecast Error (RMSFE), which penalises large deviations more heavily, and the Mean Absolute Forecast Error (MAFE), which is more robust to outliers. These are defined as follows:
\begin{equation}\label{eq:rmsfemafe}
\text{RMSFE} = \sqrt{\frac{1}{nT} \sum_{i=1}^{n} \sum_{t=1}^{T} \left( \log \hat{h}_{t}(s_i) - \log y_{t}^2(s_i) \right)^2 }, \qquad
\text{MAFE} = \frac{1}{nT} \sum_{i=1}^{n}\sum_{t=1}^{T} \left| \log \hat{h}_{t}(s_i) - \log y_{t}^2(s_i) \right|,
\end{equation}
where \(T\) is the number of forecast periods, \(\hat{h}_{t}(s_i)\) denotes the predicted conditional variance at location/node \(s_i\), and \(y_t^2(s_i)\) is the squared log return used as a proxy for realised volatility. While this proxy is standard in the literature, it can be noisy and sensitive to extremes, especially in low-frequency settings. For the predictive performance of the competing models, we adopt a one-step-ahead forecasting strategy over a fixed out-of-sample period covering the last 252 trading days (from 21 October 2023 to 20 October 2024). At each step, a one-step-ahead forecast is generated using parameters estimated once from the in-sample data. This fixed-parameter strategy is motivated by two main considerations: (i) it reflects a practical setting where computational constraints or operational requirements discourage frequent re-estimation, and (ii) it isolates the model’s internal dynamics and structure, allowing for a clean comparison of forecasting performance without interference from adaptive re-fitting.
This setup provides a rigorous test of each model’s ability to propagate volatility dynamics over time, particularly for spatiotemporal volatility models that aim to capture persistent spatial and temporal dependencies. While alternative approaches—such as rolling or expanding window re-estimation offer parameter adaptability, they may also introduce instability or obscure model-specific behaviour. By holding parameters fixed, we focus the evaluation on the model’s specification and structural assumptions.

The Diebold--Mariano (DM) test \cite{diebold1995comparing} is employed to test for equal predictive performance of each pair of models. To obtain a single cross-sectional performance measure at each time point, individual asset losses are averaged across all $N$ assets. Each model is evaluated under its best-performing spatial configuration, as identified in the previous section, thereby treating the spatial weight matrix as part of the model specification. The test is performed pairwise across spatiotemporal models using aggregated out-of-sample forecast errors. A statistically significant DM statistic indicates that one model provides superior predictive accuracy. This strategy allows for a rigorous and interpretable comparison of forecasting performance across spatially dependent volatility models.

\subsection{Data description and results}
\begin{figure}[htbp]
\centering

\begin{subfigure}[t]{0.32\textwidth}
  \includegraphics[width=\linewidth]{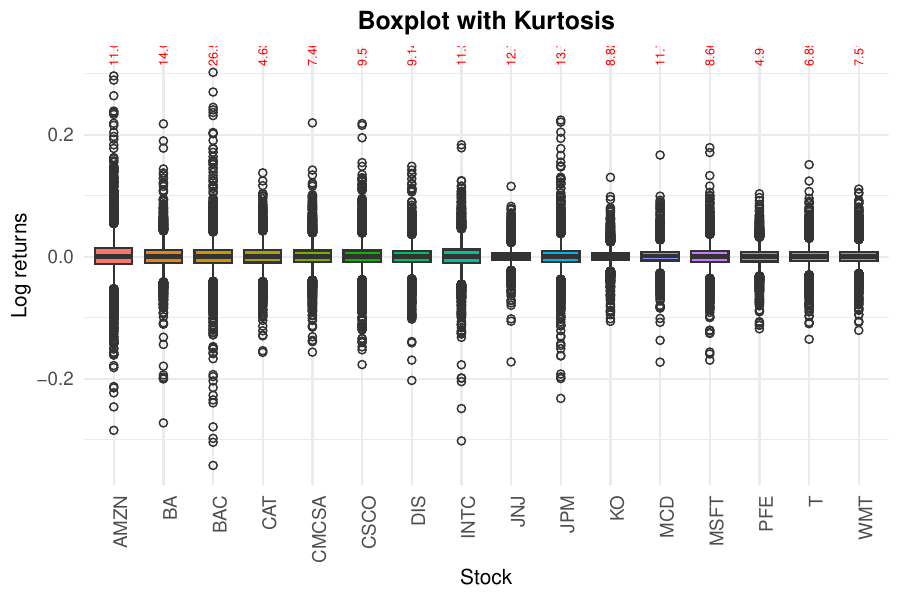}
\end{subfigure}
\begin{subfigure}[t]{0.32\textwidth}
  \includegraphics[width=\linewidth]{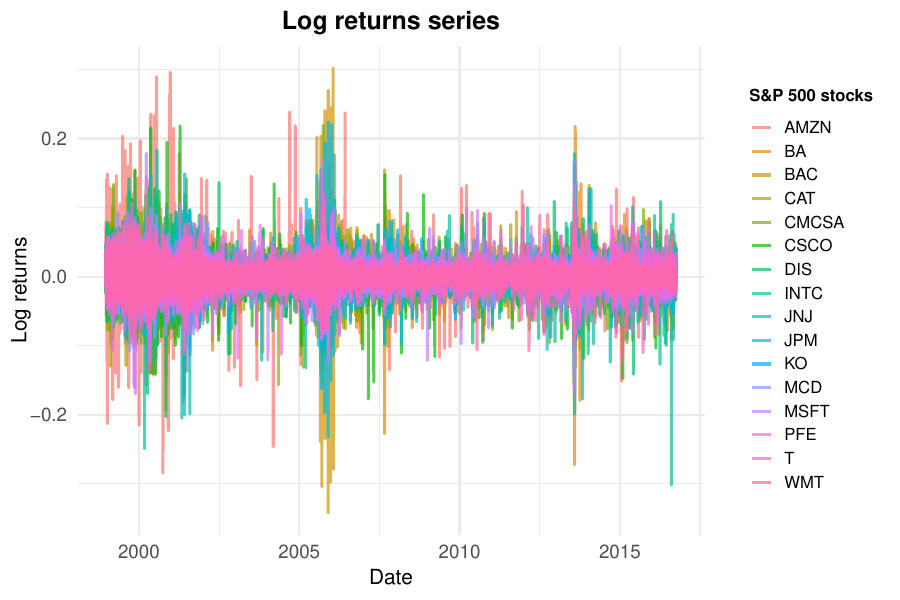}
\end{subfigure}
\begin{subfigure}[t]{0.32\textwidth}
  \includegraphics[width=\linewidth]{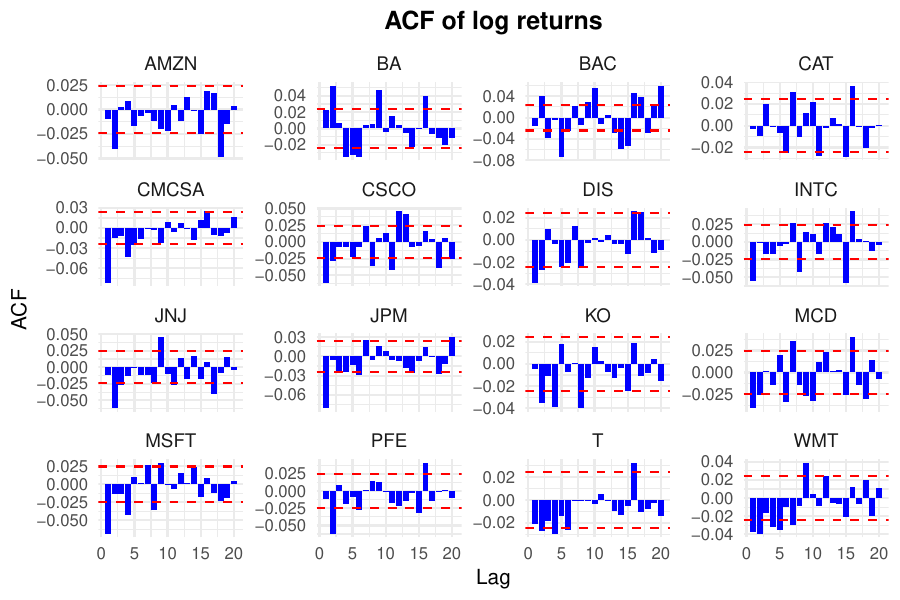}
\end{subfigure}

\vspace{1em}

\begin{subfigure}[t]{0.32\textwidth}
  \includegraphics[width=\linewidth]{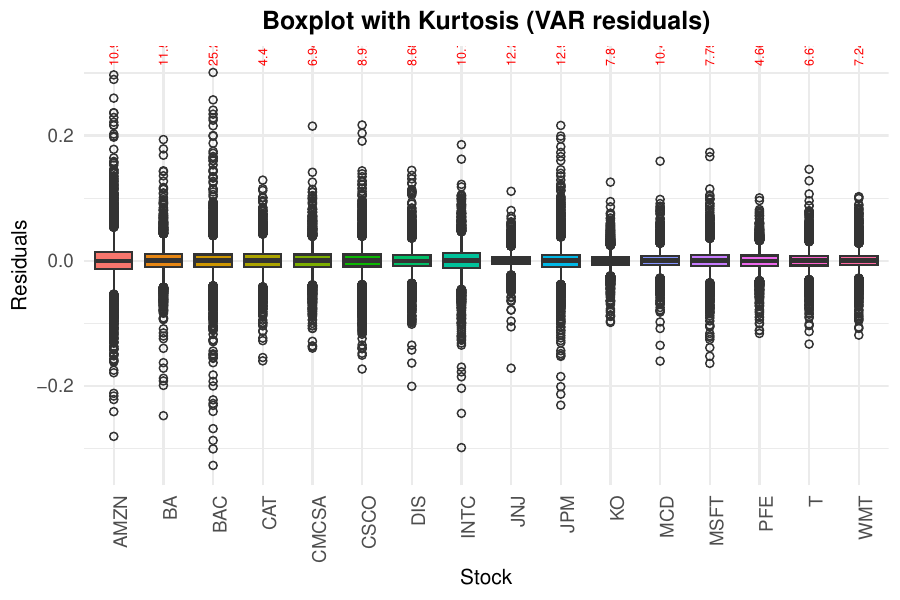}
\end{subfigure}
\begin{subfigure}[t]{0.32\textwidth}
  \includegraphics[width=\linewidth]{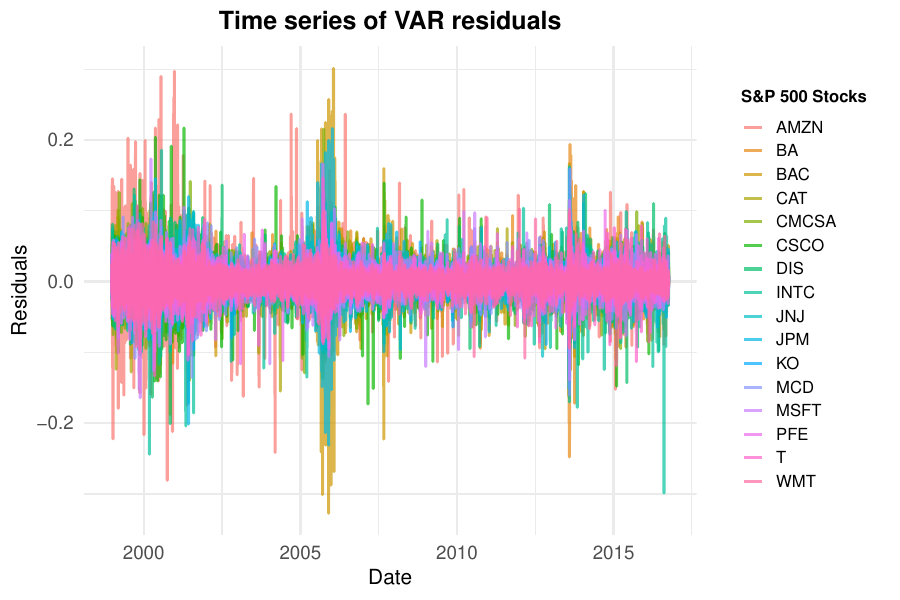}
\end{subfigure}
\begin{subfigure}[t]{0.32\textwidth}
  \includegraphics[width=\linewidth]{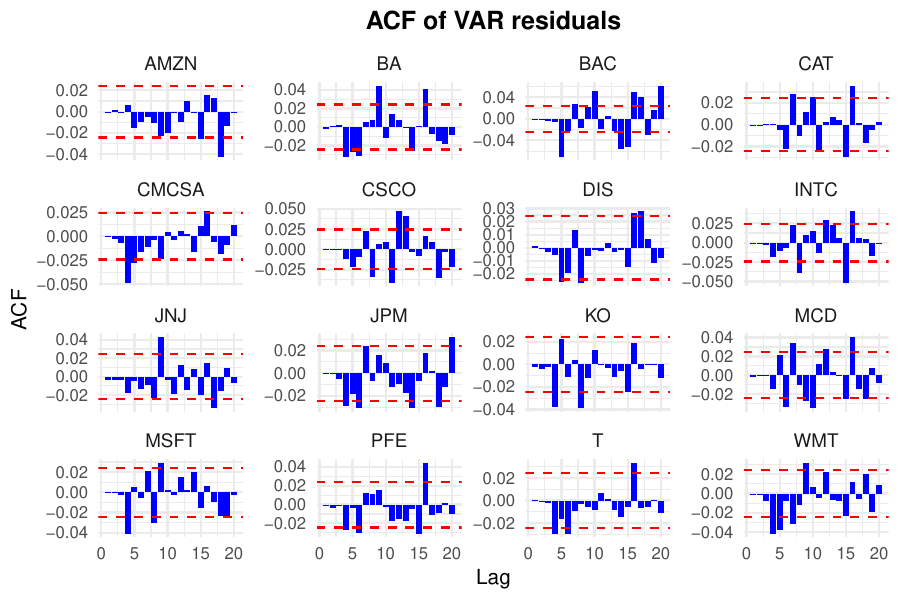}
\end{subfigure}
\vspace{1em}

\begin{subfigure}[t]{0.4\textwidth}
  \includegraphics[width=\linewidth]{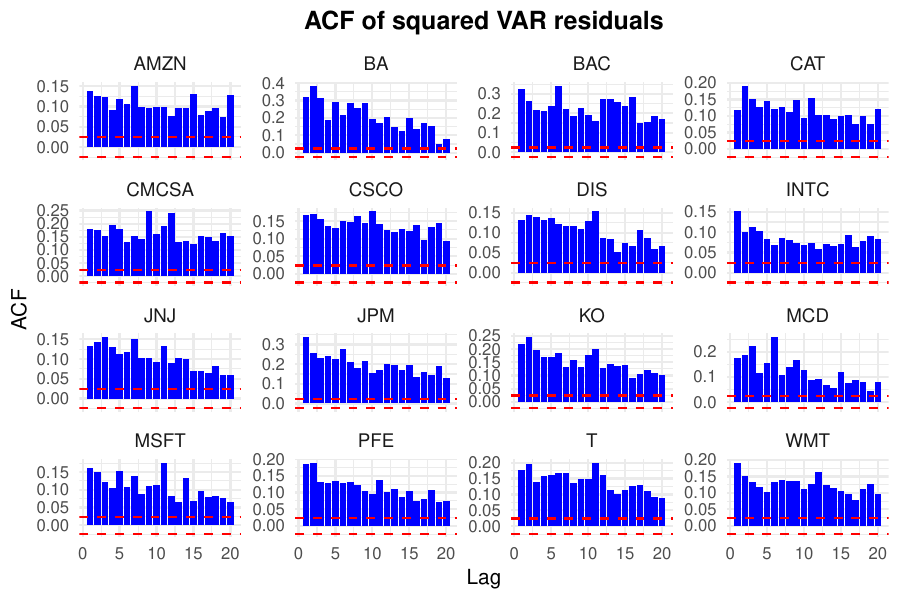}
\end{subfigure}
\begin{subfigure}[t]{0.4\textwidth}
  \includegraphics[width=\linewidth]{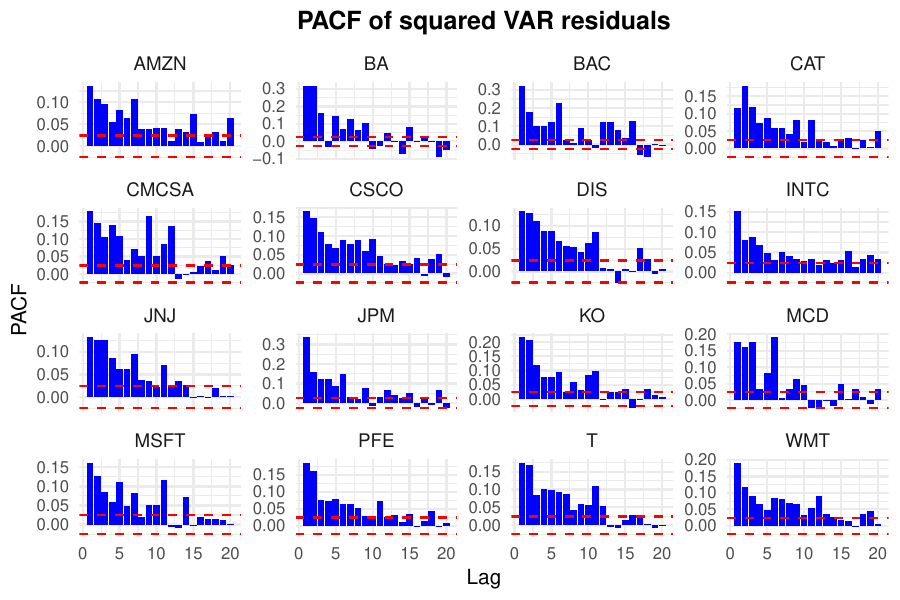}
\end{subfigure}

\vspace{1em}

\begin{subfigure}[t]{0.4\textwidth}
  \includegraphics[width=\linewidth]{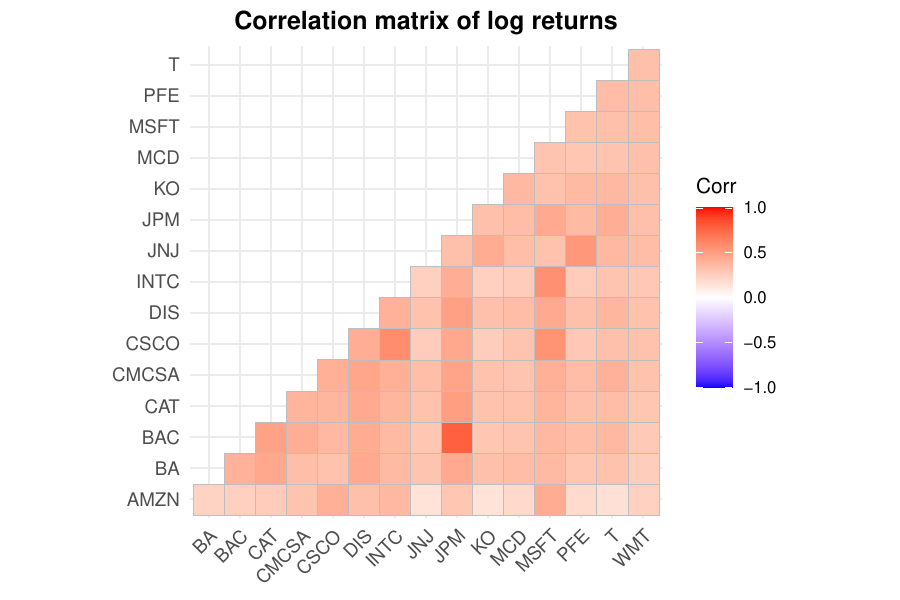}
\end{subfigure}
\begin{subfigure}[t]{0.4\textwidth}
  \includegraphics[width=\linewidth]{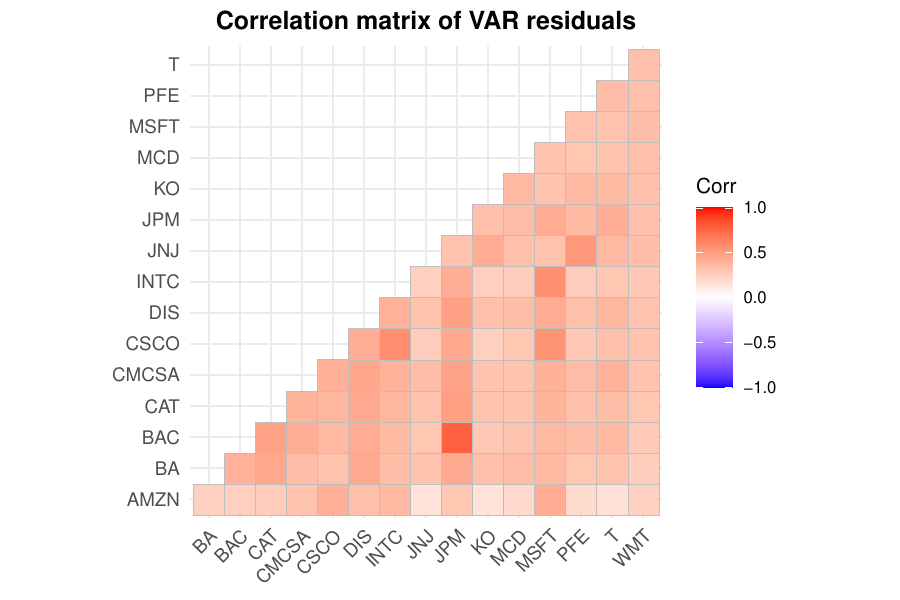}
\end{subfigure}

\caption{Diagnostics of log returns and VAR(1) residuals for 16 S$\&$P 500 stocks.}
\label{fig:diagnostics_grid}
\end{figure}

 The S$\&$P~500 is among the most widely tracked equity indices, representing 500 leading U.S. companies. In line with Modern Portfolio Theory of \cite{markowitz_portfolio_1952}, diversification is essential for mitigating risk, with empirical studies suggesting that substantial diversification benefits can be achieved with as few as 12--18 stocks \cite{reilly_investment_2018}. For large-cap portfolios, \cite{eccles_peak_2021} found marginal gains in volatility reduction beyond 15 stocks. Although \cite{zaimovic_how_2021} recommends a minimum of 30, this study selects a portfolio of at least 16 stocks to balance diversification with computational feasibility, particularly given the parameter burden in models such as DCC and BEKK.
Stock selection prioritises sectoral diversification within the S$\&$P~500. Sectors were ranked based on their prominence, defined as the proportion of years spent in the top five of index composition \cite{sather_historical_2023}. Sectors with less than $5\%$ prominence (Energy, Materials, Real Estate, Utilities) were excluded. For the remaining sectors, the number of selected stocks was scaled to their average weight in the index. This resulted in a selection of 16 stocks spanning 22 December 1998 to 20 October 2024, with 6,244 daily training observations and 252 for testing.
EGARCH diagnostics indicate that $94.59\%$ of the series exhibit leptokurtosis and $91.89\%$ display significant asymmetry, highlighting the non-Gaussian features of the data. Figure \ref{fig:diagnostics_grid} presents diagnostic plots for the log returns and residuals from a fitted VAR(1) model for the 16  S$\&$P~500 stocks. The boxplots and time series plots confirm key stylised facts, including heavy tails, volatility clustering, and episodes of heightened turbulence. While the VAR model removes much of the linear mean dependence—evidenced by residuals centred around zero with reduced dispersion—diagnostics reveal persistent conditional heteroskedasticity and strong cross-sectional dependence. Autocorrelation functions show weak serial dependence, supporting the limited linear predictability of returns. However, the residuals still display volatility clustering and existing correlations, indicating the presence of latent spatiotemporal structures. These findings motivate the use of multivariate and spatiotemporal GARCH-type models to better capture the dynamics of financial market volatility. Because the evaluation focuses on forecasted variances, only the diagonal elements of the conditional covariance matrices were retained when assessing forecasting performance, even for models that specify the full covariance structure.

\newpage

\begin{center}
\small
\begin{longtable}{l c c c  c c c}
   
    \label{tab:model_performance}\\
    \hline
    \textbf{Model} & $k$ & \textbf{Time}  & \shortstack{\textbf{Weight} \\ \textbf{Matrix}} & \textbf{BIC} & \textbf{RMSFE} & \textbf{MAFE} \\
    \hline
    \endfirsthead

    \hline
    \textbf{Model} & $k$ & \textbf{Time} & \shortstack{\textbf{Weight} \\ \textbf{Matrix}} & \textbf{BIC} & \textbf{RMSFE} & \textbf{MAFE} \\
    \hline
    \endhead

    \hline
    \multicolumn{1}{r}{\textit{Continued on the next page}} \\
    \hline
    \endfoot

    \hline
    \endlastfoot

    \multirow{1}{*}{DCC} & 170 & 0.13 h & - &  \textbf{-572707.5}  & 2.908 & 2.100 \\

    \hline
    \multirow{1}{*}{Diagonal BEKK symmetric} & 168 & 30.3 h & -  & -570981.2 & 2.923 &    2.112 \\
    \hline
    \multirow{1}{*}{Diagonal BEKK asymmetric} & 184 &  64.4 h & -  &-571236.7 & 2.908 &   2.099  \\

    \hline

    Spatial GARCH-X & 64 & 2.36 h
    & $\textbf{W}^{Eucli}$   & -538710.5 &  2.883 & 2.084  \\
 &   &    & $\textbf{W}^{Corr}$  &  -538702.5 &   2.883 & 2.084 \\
 &   &   & $\textbf{W}^{Picco}$  &  -538703.8 & 2.883 & 2.083 \\
&   &   & $\textbf{W}^{E-G}$ &  -538605.5 &  2.885 & 2.086    \\
   & &   & $\textbf{W}^{C-G}$ &  -538603.0 & 2.886 & 2.086   \\
 &  &  & $\textbf{W}^{P-G}$ &  -538605.5 & 2.885 & 2.085  \\
 &  &    & $\textbf{W}^{E-NN}$ &  -538826.2 & 2.884 & 2.083  \\
  & &   & $\textbf{W}^{C-NN}$ &  -538825.3 & 2.884 & 2.084   \\
 &  &  & $\textbf{W}^{P-NN}$ &  -538811.0 & 2.883 & 2.083   \\
 &   & & $\textbf{W}^{Spill}$ &  -538816.2 &  2.883 & 2.084   \\

    \hline

    Dynamic Log-ARCH & 17 & \textbf{0.041 h} & $\textbf{W}^{Eucli}$  & -463484.5 &  2.627 & 1.985    \\
 &  &   & $\textbf{W}^{Corr}$  & -461893.5  & 2.616 &  1.973     \\
  & &  & $\textbf{W}^{Picco}$  & -465524.5 & 2.618 &  1.975  \\
 &  &      & $\textbf{W}^{E-G}$ & -479089.2  & 2.550 &  1.911   \\
 & &   & $\textbf{W}^{C-G}$ &  -478544.8 & \textbf{2.543} &  \textbf{1.902}   \\
 & &   & $\textbf{W}^{P-G}$ & - 481341.6 &  2.591 &  1.945  \\
 & &    & $\textbf{W}^{E-NN}$ & -472596.0  & 2.975 &  2.280      \\
 & &   & $\textbf{W}^{C-NN}$ & -481438.9 &  2.664 &  1.984     \\
 & &   & $\textbf{W}^{P-NN}$ & -468878.7  &  2.748 & 2.070     \\
 &  &  & $\textbf{W}^{Spill}$ &  -463867.3 &  2.643 &  2.001     \\

     \hline

Circular spatiotemporal GARCH & 6 & 0.31 h & $\textbf{W}^{Eucli}$  &  -517017.5 & 2.951 & 2.142      \\
 & & & $\textbf{W}^{Corr}$  &  -524303.5 & 2.873 & 2.069    \\
 & & & $\textbf{W}^{Picco}$  &  -516562.7 & 2.951 & 2.141    \\
 & &  & $\textbf{W}^{E-G}$ &  -507642.6 & 3.111 & 2.293    \\
 & & & $\textbf{W}^{C-G}$ & -507361.9 & 3.115 & 2.297   \\
 & & & $\textbf{W}^{P-G}$ &  -507344.6 & 3.109 & 2.291     \\
 & & & $\textbf{W}^{E-NN}$ &  -523688.8 & 3.179 & 2.360     \\
 & & & $\textbf{W}^{C-NN}$ &  -525687.0 &2.803 & 2.011      \\
 & & & $\textbf{W}^{P-NN}$ &  -523116.3 & 2.924 & 2.115     \\
 & & & $\textbf{W}^{Spill}$ &  -523861.7 & 2.900 & 2.093     \\
\hline

Spatiotemporal EGARCH & 7 & 11.8 h & $\textbf{W}^{Eucli}$  &  -523367.2 &    3.292 & 2.467    \\
 & &  & $\textbf{W}^{Corr}$  &  -536222.4 & 4.821 & 3.485   \\
 & & & $\textbf{W}^{Picco}$  &  -488616.1 & 3.883 & 2.972     \\
 &   &  & $\textbf{W}^{E-G}$ &  -488801.2 &  3.292 & 2.467   \\
 & & & $\textbf{W}^{C-G}$ &  -536420.8 &   3.754  & 2.806  \\
 & & & $\textbf{W}^{P-G}$ &  -485491.4 &    4.740 & 3.440 \\
 & & & $\textbf{W}^{E-NN}$ &  -537771.6 &   4.835 & 3.500   \\
 & & & $\textbf{W}^{C-NN}$ &  -533900.6 &   4.386 & 3.208  \\
 & & & $\textbf{W}^{P-NN}$ &  - 526937.5 & 3.460 & 2.597  \\
 & & & $\textbf{W}^{Spill}$ &  - 521640.6 & 3.486 & 2.620   \\
\hline

Proximity-structured BEKK & 21 & 51.4 h & $\textbf{W}^{Eucli}$  &   -309661.9 & 5.715 & 5.182    \\
 & & & $\textbf{W}^{Corr}$  &  -309683.4 &  5.715 & 5.182  \\
 & & & $\textbf{W}^{Picco}$  &  -309651.0 &  5.715 & 5.182  \\
 &  &   & $\textbf{W}^{E-G}$ &  -309731.7 &  5.721 & 5.189  \\
 & & & $\textbf{W}^{C-G}$ &  -309742.2 &  5.721 & 5.189  \\
 & & & $\textbf{W}^{P-G}$ & -309735.7 &  5.722 & 5.190  \\
 & & & $\textbf{W}^{E-NN}$ &  -309552.4 &  5.717 & 5.184  \\
 & & & $\textbf{W}^{C-NN}$ &  -309662.7 &  5.717 & 5.184  \\
 & & & $\textbf{W}^{P-NN}$ &  -309605.9 &  5.718 & 5.185  \\
 & & & $\textbf{W}^{Spill}$ &  -309624.6 &  5.714 & 5.182  \\
\hline

 \caption{Model performance metrics across weight matrices. Bayesian Information Criterion (BIC), Root Mean Squared Forecast Error (RMSFE), and Mean Absolute Forecast Error (MAFE) for models estimated using different spatial weight matrices. The best value in each column is printed in bold. $k$ denotes the number of estimated parameters for each model.} 
\end{longtable}
\end{center}

Table~\ref{tab:model_performance} summarises the forecasting and in-sample performance of all models across the spatial weight matrices introduced in Section~\ref{sec3}. Several clear patterns emerge.
First, the Dynamic Spatiotemporal ARCH model consistently achieves the best out-of-sample forecast accuracy, with the lowest RMSFE and MAFE across nearly all network specifications. Its strongest performance is observed under Granger-filtered matrices such as $\mathbf{W}^{C\text{-}G}$ (RMSFE = $2.543$, MAFE = $1.902$). Despite its very low dimensionality, the model effectively captures spatiotemporal volatility persistence, and its GMM-based estimation avoids the numerical instabilities that affect likelihood-based approaches.
Second, among the likelihood-based models, the DCC specification achieves the lowest BIC across most spatial configurations, indicating a favourable balance between model fit and parameter complexity. Its forecasting performance is also strong, slightly exceeding both the symmetric and asymmetric diagonal BEKK models. Although introducing asymmetry into the BEKK model leads to minor improvements in MAFE, the overall accuracy remains comparable to that of the symmetric version.
Third, the Spatial GARCH-X and Circular Spatiotemporal GARCH models perform competitively, particularly when Granger-filtered or nearest-neighbour matrices are used. For example, the Circular STGARCH specification achieves its best results under $\mathbf{W}^{C\text{-}NN}$, suggesting that parsimonious spatial structures can enhance numerical stability and improve predictive accuracy.

By contrast, the Spatiotemporal EGARCH model performs poorly across all spatial configurations, with substantially higher RMSFE and MAFE values. This weak performance is due to numerical instability in the nonlinear inversion step of the spatiotemporal recursion, which is highly sensitive to the random initial values used during optimisation. 
Finally, the Proximity-Structured multivariate model yields the weakest performance overall, with the highest forecast errors and the longest computation times. Although faster implementations exist when closed-form derivatives are available, as in the BEKK package developed by \cite{fulle2024bekks}, these techniques do not apply to the structural constraints of the proximity-based formulation used here. Computations were performed on a machine with four Intel Xeon Platinum 8276L CPUs (each with 28 cores, for a total of 224 logical cores) and 1 TB of RAM.

The results highlight the substantial value of incorporating spatial structure into volatility modelling. Models that successfully exploit spatial information, particularly through filtered or nearest-neighbour matrices, consistently outperform traditional multivariate GARCH frameworks. The strong performance of the Dynamic Log-ARCH model further underscores the numerical advantages of GMM-based estimation over full maximum-likelihood methods in high-dimensional or spatially dependent volatility environments. The choice of spatial weight matrix therefore emerges as a central modelling component, strongly shaping both estimation robustness and forecasting accuracy in financial network applications.

\begin{table}[h!]
\tiny
\centering
\begin{tabular}{ccccccccccc}
  \toprule
Model & Mean Loss & Rank & BEKK & ABEKK & DCC & DSTARCH  & SpGARCH-X & STE-GARCH & STBEKK & STGARCH \\ 
  \midrule
DSTARCH & 6.514 & 1 & \textbf{<0.001} & \textbf{<0.001} & \textbf{<0.001} & -- & \textbf{<0.001} & \textbf{<0.001} & \textbf{<0.001} & \textbf{<0.001} \\ 
  STGARCH & 7.859 & 2 & \textbf{<0.001} & \textbf{<0.001} & \textbf{<0.001} & \textbf{<0.001} & \textbf{<0.001} & \textbf{<0.001} & \textbf{<0.001} & -- \\ 
  SpGARCH-X & 8.345 & 3 & \textbf{<0.001} & \textbf{<0.001} & \textbf{0.007} & \textbf{<0.001} & -- & \textbf{<0.001} & \textbf{<0.001} & \textbf{<0.001} \\ 
  DCC & 8.491 & 4 & 0.105 & 0.979 & -- & \textbf{<0.001} & \textbf{0.007} & \textbf{<0.001} & \textbf{<0.001} & \textbf{<0.001} \\ 
  ABEKK & 8.504 & 5 & \textbf{<0.001} & -- & 0.979 & \textbf{<0.001} & \textbf{<0.001} & \textbf{<0.001} & \textbf{<0.001} & \textbf{<0.001} \\ 
  BEKK & 8.591 & 6 & -- & \textbf{<0.001} & 0.105 & \textbf{<0.001} & \textbf{<0.001} & \textbf{<0.001} & \textbf{<0.001} & \textbf{<0.001} \\ 
  STEGARCH & 11.973 & 7 & \textbf{<0.001} & \textbf{<0.001} & \textbf{<0.001} & \textbf{<0.001} & \textbf{<0.001} & -- & \textbf{<0.001} & \textbf{<0.001} \\ 
  STBEKK & 32.658 & 8 & \textbf{<0.001} & \textbf{<0.001} & \textbf{<0.001} & \textbf{<0.001} & \textbf{<0.001} & \textbf{<0.001} & -- & \textbf{<0.001} \\ 
   \bottomrule
\end{tabular}
\caption{Pairwise Diebold--Mariano (DM) test p-values for equality of predictive accuracy. 
              Smaller p-values indicate significant differences. 
              The two leftmost columns report each model’s average loss and rank (lower = better). 
              Bold entries denote strong significance (p < 0.01); italics denote moderate significance (p < 0.05).} \label{tab:dm-summary}
\end{table}

Table~\ref{tab:dm-summary} reports pairwise Diebold--Mariano (DM) $p$-values for equality of predictive accuracy alongside each model's mean loss and rank. The ordering by mean loss reveals a clear hierarchy: DSTARCH attains the lowest average loss (6.51; Rank~1), followed by STGARCH (7.86; Rank~2). spGARCH-X (8.34; Rank~3) and the DCC/BEKK family (8.49--8.59; Ranks~4--6) form a mid-performing cluster, whereas STEGARCH (11.97; Rank~7) and especially the proximity-structured BEKK model (32.66; Rank~8) exhibit substantially larger losses.
The DM $p$-values corroborate these differences. Nearly all cross-model comparisons reject equal predictive accuracy at conventional levels (typically $p<0.001$), indicating materially different forecast performance across specifications. Two notable exceptions align with the ranking: DCC vs.\ ABEKK yields $p\approx 0.979$, and DCC vs.\ BEKK yields $p\approx 0.105$, implying statistically indistinguishable accuracy between these BEKK/DCC variants. In contrast, spGARCH-X vs.\ DCC shows $p\approx 0.007$, indicating a significant advantage for spGARCH-X over DCC. The STEGARCH and STBEKK differ significantly from all competitors ($p<0.01$ throughout), which is consistent with their high mean losses and suggests inferior out-of-sample volatility forecasts.
\noindent Overall, the evidence supports the conclusion that DSTARCH is the top-performing specification, STGARCH is competitive but slightly weaker, SpGARCH-X and DCC/BEKK constitute a middle tier with some pairwise equivalences, and proximity-based formulations underperform markedly.

\section{Conclusion}\label{sec:conclusion}

This study evaluated and compared the forecasting performance of several spatiotemporal volatility models within a financial network framework. The goal was to assess whether incorporating spatial dependence enhances volatility forecasting relative to traditional multivariate GARCH benchmarks. Using daily returns from 16 sectorally diversified S$\&$P~500 stocks, each model was estimated once on a training period and used to generate recursive one-step-ahead forecasts on a fixed out-of-sample window, ensuring a consistent comparison across specifications.

The results show that spatial dependence can significantly improve forecasting accuracy. Among all models, the Dynamic Log-ARCH delivered the lowest RMSFE and MAFE across weight matrices, outperforming both classical and spatiotemporal alternatives. The Spatiotemporal EGARCH, although designed to capture key stylised facts such as asymmetry and nonlinear spillovers, produced weaker forecasts due to numerical instability in its inversion step, which made optimisation highly sensitive to initial values. The stationary and standard Spatial GARCH models offered moderate accuracy while maintaining relatively simple parameter structures.

The choice of spatial weight matrix played a central role in model performance. Correlation-based networks consistently outperformed Euclidean and log-ARCH-based matrices, and the use of Granger-filtered weights further improved accuracy by isolating causal dependence structures. In particular, the Dynamic Log-ARCH combined with the Granger-filtered correlation matrix ($\mathbf{W}^{C-G}$) yielded the best overall forecast accuracy, indicating that economic connectivity rather than geometric proximity drives volatility co-movement among assets. In terms of in-sample fit, however, classical multivariate models such as BEKK and DCC achieved the lowest BIC values.

Pairwise Diebold–Mariano tests confirmed statistically significant differences in predictive accuracy (mostly $p < 0.01$), supporting the dominance of the Dynamic Spatiotemporal ARCH model. Its strong performance appears closely tied to the numerical stability of its GMM-based estimation, which avoids many of the convergence issues encountered in likelihood-based spatiotemporal models. Taken together, the findings suggest that spatial econometric methods—particularly those based on causally filtered correlation networks—provide a powerful and interpretable framework for modelling volatility in interconnected markets.

Future research could incorporate intraday realised volatility measures to achieve more refined forecast evaluations and extend the analysis to larger financial systems where sparsity and regularisation are essential for stability. Another promising direction is to allow the spatial weight matrix to evolve over time or to estimate it directly from the data, enabling models to capture structural shifts in financial connectivity. Given the strong empirical performance of causally filtered correlation matrices, exploring regime-dependent or data-driven network structures may further enhance volatility modelling.

Overall, the evidence indicates that spatially structured volatility models, despite their parsimonious parameterisation, can deliver forecasting performance that rivals high-dimensional multivariate models. In environments where classical specifications become computationally burdensome or unstable, spatiotemporal volatility models offer a flexible, interpretable, and efficient alternative for forecasting volatility in complex financial networks.

\section*{Acknowledgement}
We gratefully acknowledge funding by the Deutsche Forschungsgemeinschaft (DFG, German Research Foundation) project number 501539976.

\newpage

\bibliography{reference}

\newpage

\begin{appendix}
\section{Appendix}

\subsection{Exploratory analysis}
The full set of scripts used to produce the empirical results and forecasting evaluations reported in this paper is available at
\url{https://github.com/ariane237/Spatiotemporal_Volatility_Comparison}.

\begin{table}[h!]
\centering
\caption{Selected S\&P 500 stocks with tickers, company names, and sector classifications.}
\label{tab:ticker_sector_merged}
\begin{tabular}{lll}
\toprule
\textbf{Sector} & \textbf{Ticker} & \textbf{Company Name} \\
\midrule
\multirow{2}{*}{Consumer Discretionary}
  & AMZN  & Amazon.com Inc. \\
  & MCD   & McDonald's Corporation \\
\midrule
\multirow{3}{*}{Communication Services}
  & DIS   & The Walt Disney Company \\
  & T     & AT\&T Inc. \\
  & CMCSA & Comcast Corporation \\
\midrule
\multirow{2}{*}{Consumer Staples}
  & WMT   & Walmart Inc. \\
  & KO    & The Coca-Cola Company \\
\midrule
\multirow{2}{*}{Financials}
  & JPM   & JPMorgan Chase \& Co. \\
  & BAC   & Bank of America Corporation \\
\midrule
\multirow{2}{*}{Healthcare}
  & JNJ   & Johnson \& Johnson \\
  & PFE   & Pfizer Inc. \\
\midrule
\multirow{2}{*}{Industrials}
  & BA    & The Boeing Company \\
  & CAT   & Caterpillar Inc. \\
\midrule
\multirow{3}{*}{Technology}
  & INTC  & Intel Corporation \\
  & MSFT  & Microsoft Corporation \\
  & CSCO  & Cisco Systems Inc. \\
\bottomrule
\end{tabular}
\end{table}

\begin{longtable}{lccccccc}
\caption{Summary statistics of daily log-returns. The table reports statistics on the univariate distributions (mean, standard deviation, minimum, maximum, skewness, kurtosis, and ARCH test results).}
\label{table:ticker_statistics} \\
\hline
\textbf{Ticker} & \textbf{Mean} & \textbf{Std. Dev.} & \textbf{Min} & \textbf{Max} & \textbf{Skewness} & \textbf{Kurtosis} & \textbf{ARCH Test (p-value)} \\
\hline
\endfirsthead
\hline
\textbf{Ticker} & \textbf{Mean} & \textbf{Std. Dev.} & \textbf{Min} & \textbf{Max} & \textbf{Skewness} & \textbf{Kurtosis} & \textbf{ARCH Test (p-value)} \\
\hline
\endhead
\hline
\multicolumn{3}{r}{\textit{Continued on the next page}} \\
\hline
\endfoot
\hline
\endlastfoot
AMZN  & 0.001 & 0.033 & -0.285 & 0.296 & 0.395  & 14.026 & $<$0.001 \\
        BA    & 0.000 & 0.022 & -0.272 & 0.218 & -0.387 & 17.064 & $<$0.001 \\
        BAC   & 0.000 & 0.027 & -0.342 & 0.302 & -0.303 & 29.510 & $<$0.001 \\
        CAT   & 0.001 & 0.020 & -0.157 & 0.137 & -0.187 & 7.632  & $<$0.001 \\
        CMCSA & 0.000 & 0.021 & -0.156 & 0.219 & 0.072  & 10.462 & $<$0.001 \\
        CSCO  & 0.000 & 0.023 & -0.177 & 0.218 & 0.042  & 12.502 & $<$0.001 \\
        DIS   & 0.000 & 0.019 & -0.203 & 0.148 & -0.046 & 12.145 & $<$0.001 \\
        INTC  & 0.000 & 0.024 & -0.302 & 0.183 & -0.742 & 14.373 & $<$0.001 \\
        JNJ   & 0.000 & 0.012 & -0.173 & 0.115 & -0.362 & 15.750 & $<$0.001 \\
        JPM   & 0.000 & 0.024 & -0.232 & 0.224 & 0.217  & 16.796 & $<$0.001 \\
        KO    & 0.000 & 0.013 & -0.106 & 0.130 & -0.158 & 11.827 & $<$0.001 \\
        MCD   & 0.000 & 0.015 & -0.173 & 0.167 & -0.112 & 14.714 & $<$0.001 \\
        MSFT  & 0.000 & 0.019 & -0.170 & 0.179 & -0.115 & 11.660 & $<$0.001 \\
        PFE   & 0.000 & 0.016 & -0.118 & 0.103 & -0.176 & 7.904  & $<$0.001 \\
        T     & 0.000 & 0.016 & -0.135 & 0.151 & -0.041 & 9.894  & $<$0.001 \\
        WMT   & 0.000 & 0.015 & -0.121 & 0.111 & 0.079  & 10.515 & $<$0.001 \\
\end{longtable}

\begin{figure}[h!]
    \centering
      \caption{Network visualisations constructed from 10 different spatial weight matrices and a sector-based legend. Networks are based on Euclidean, correlation, and Piccolo dissimilarities, their Granger-filtered and 5-nearest-neighbour versions, and one EGARCH-based volatility spillover network. Nodes represent S$\&$P 500 stocks and are colored by sector; edge darkness reflects connection strength.}
    \label{fig:network_grid}

    \begin{minipage}[t]{0.31\linewidth}
        \centering
        \includegraphics[width=\linewidth]{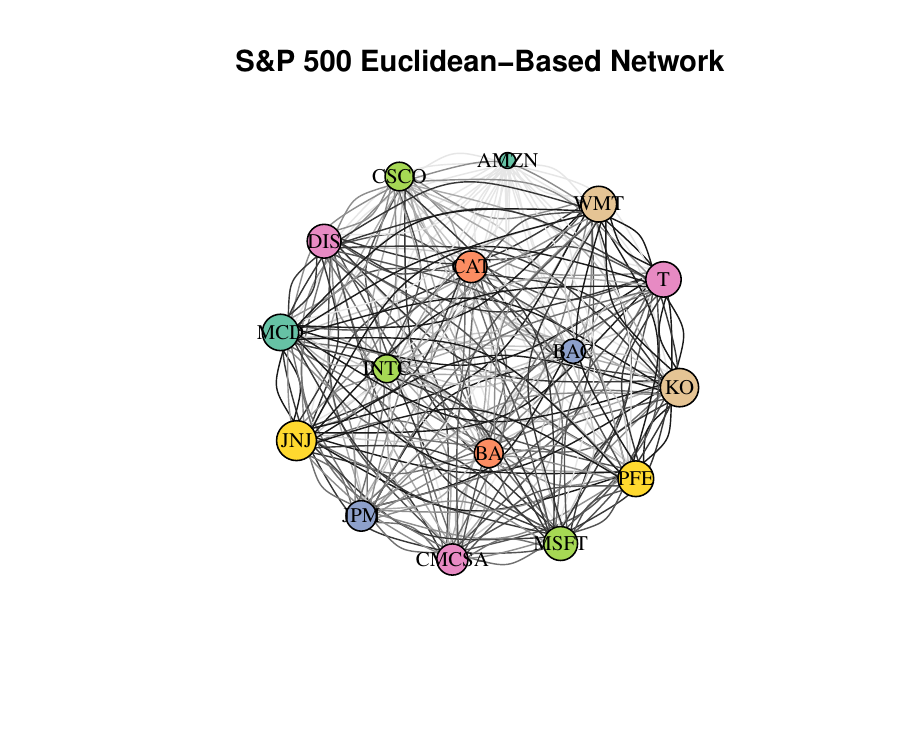}
    \end{minipage}%
    \hfill
    \begin{minipage}[t]{0.31\linewidth}
        \centering
        \includegraphics[width=\linewidth]{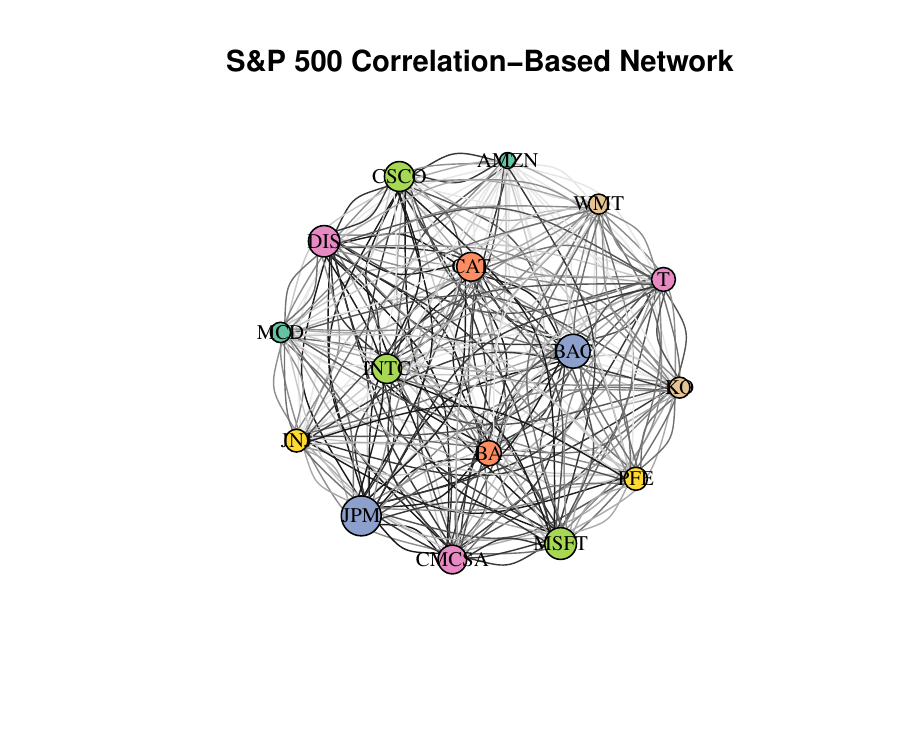}
    \end{minipage}%
    \hfill
    \begin{minipage}[t]{0.31\linewidth}
        \centering
        \includegraphics[width=\linewidth]{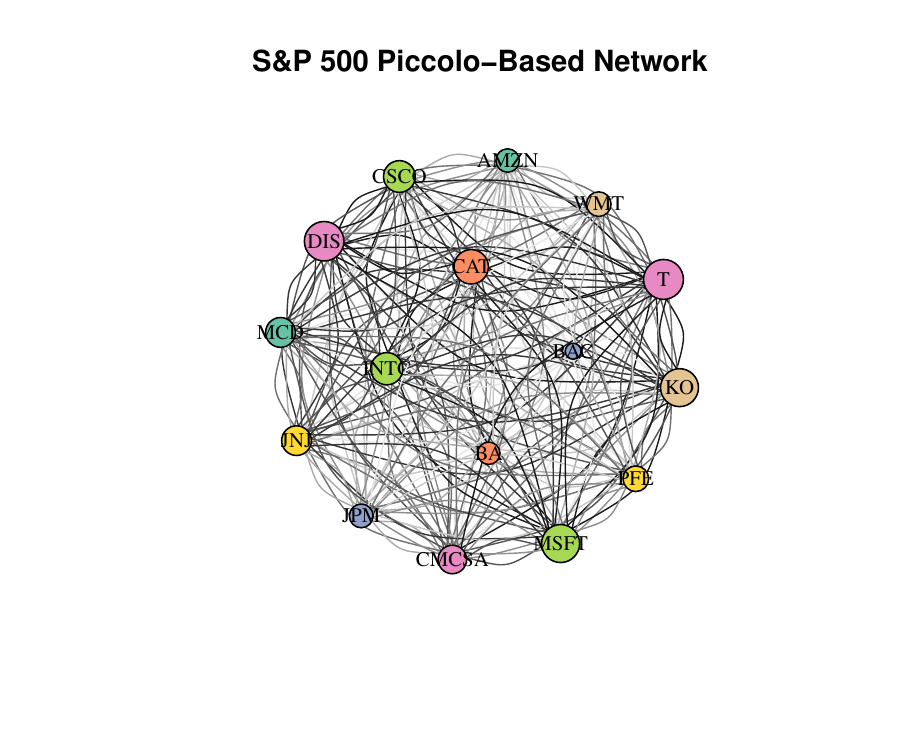}
    \end{minipage}

    \vspace{1em}
    \begin{minipage}[t]{0.31\linewidth}
        \centering
        \includegraphics[width=\linewidth]{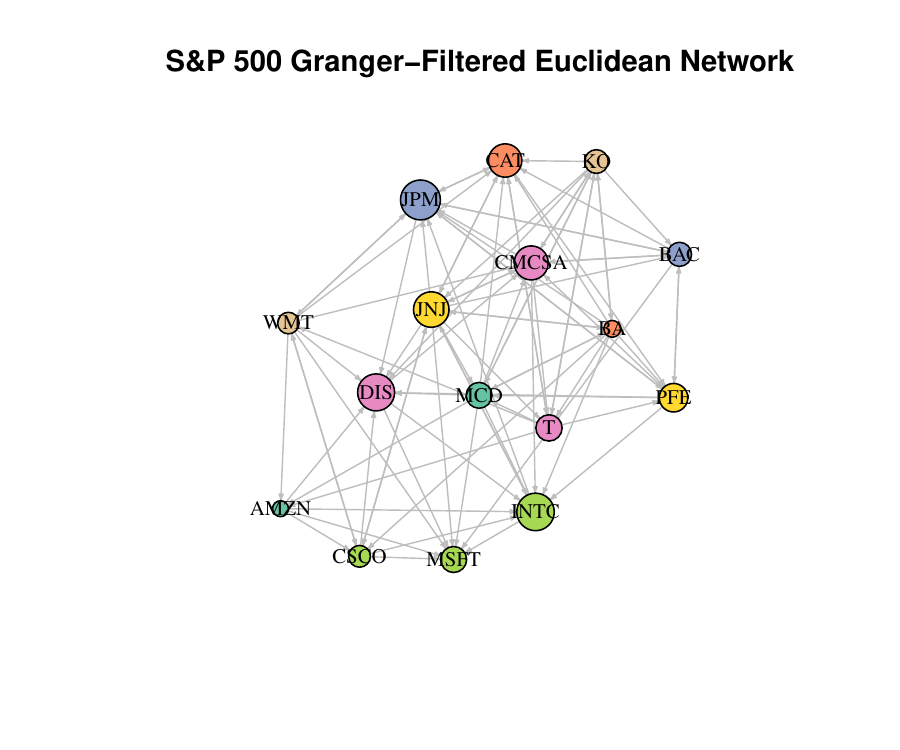}
    \end{minipage}%
    \hfill
    \begin{minipage}[t]{0.31\linewidth}
        \centering
        \includegraphics[width=\linewidth]{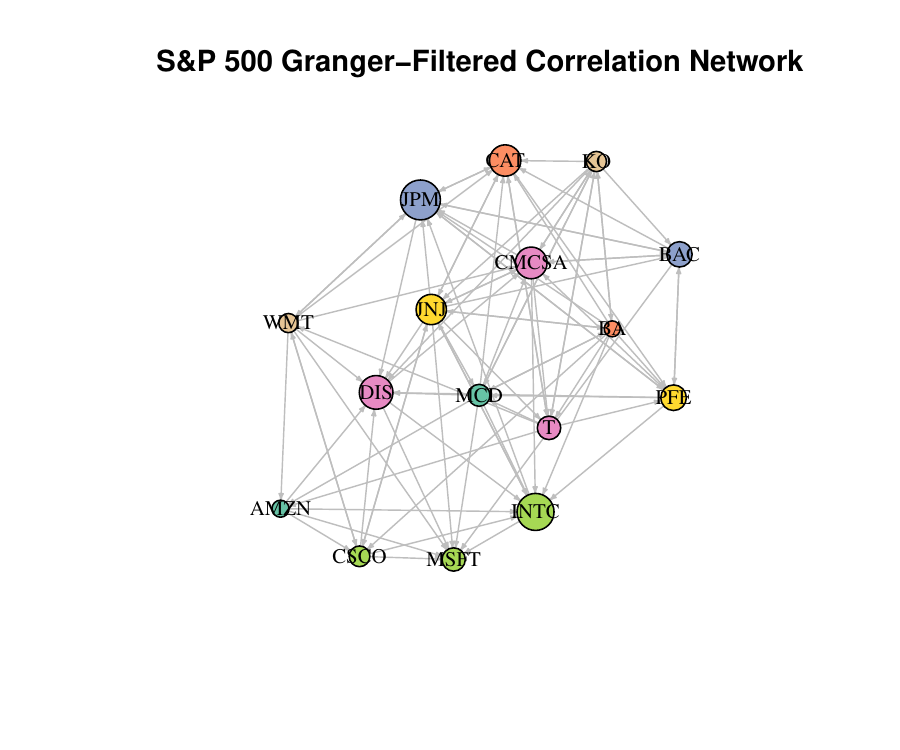}
    \end{minipage}%
    \hfill
    \begin{minipage}[t]{0.31\linewidth}
        \centering
        \includegraphics[width=\linewidth]{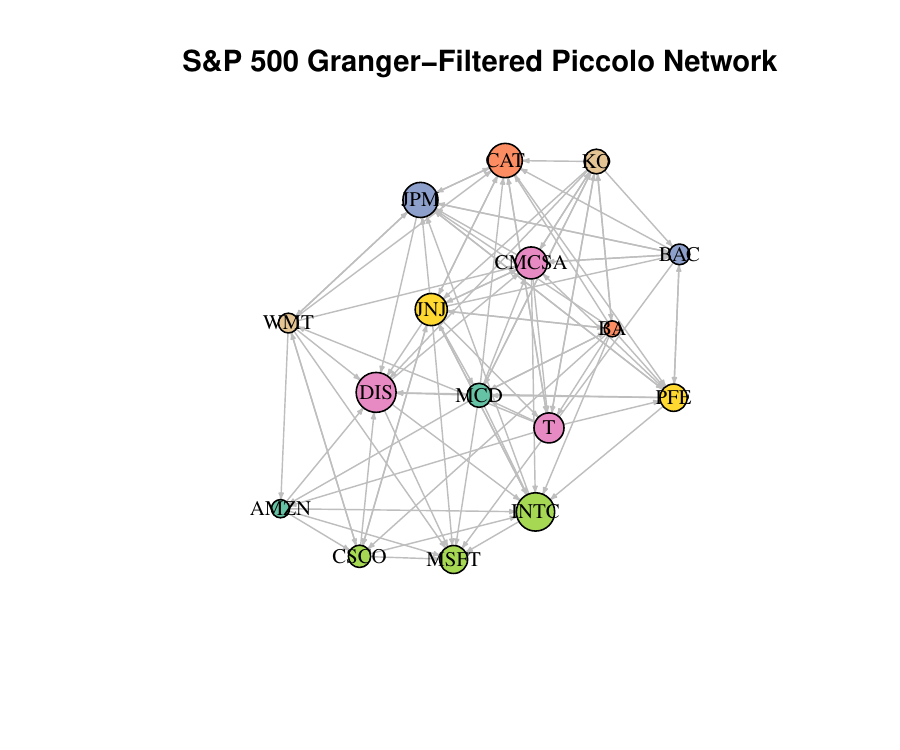}
    \end{minipage}

    \vspace{1em}
    \begin{minipage}[t]{0.31\linewidth}
        \centering
        \includegraphics[width=\linewidth]{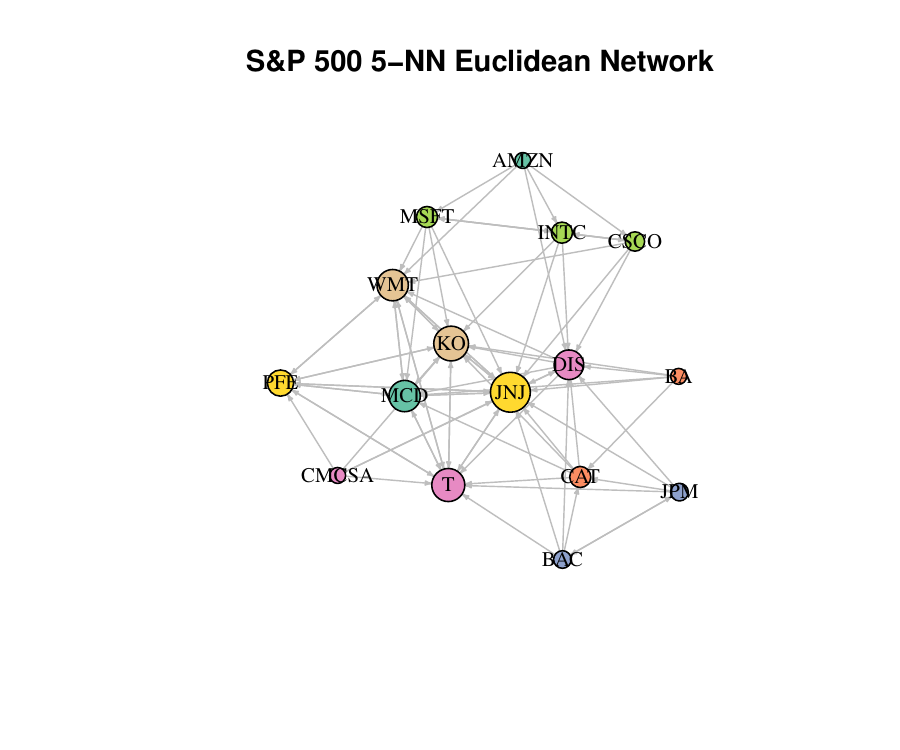}
    \end{minipage}%
    \hfill
    \begin{minipage}[t]{0.31\linewidth}
        \centering
        \includegraphics[width=\linewidth]{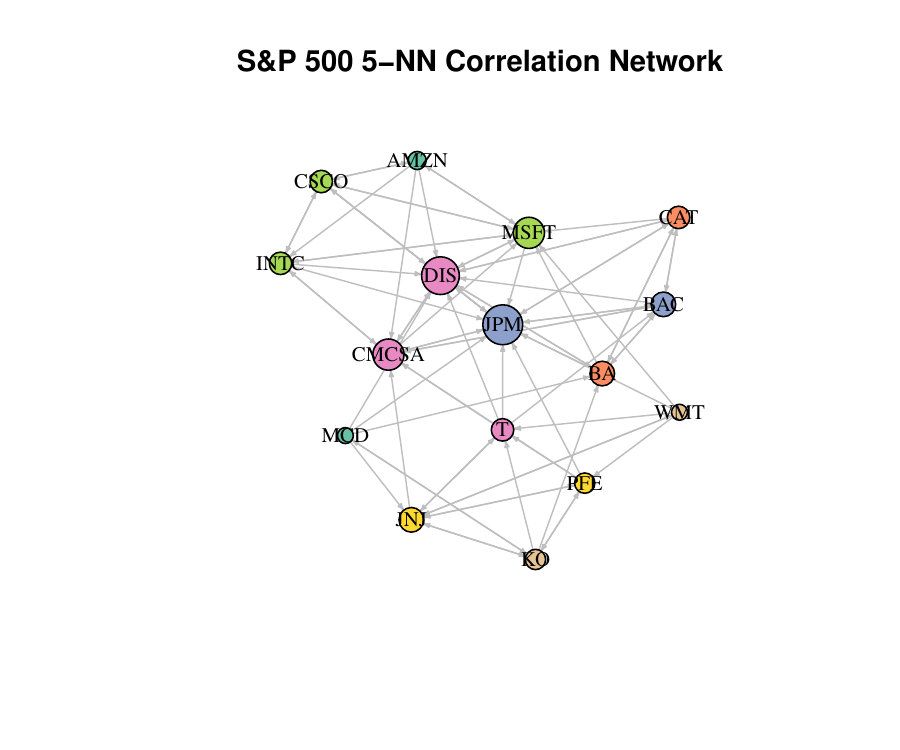}
        
    \end{minipage}%
    \hfill
    \begin{minipage}[t]{0.31\linewidth}
        \centering
        \includegraphics[width=\linewidth]{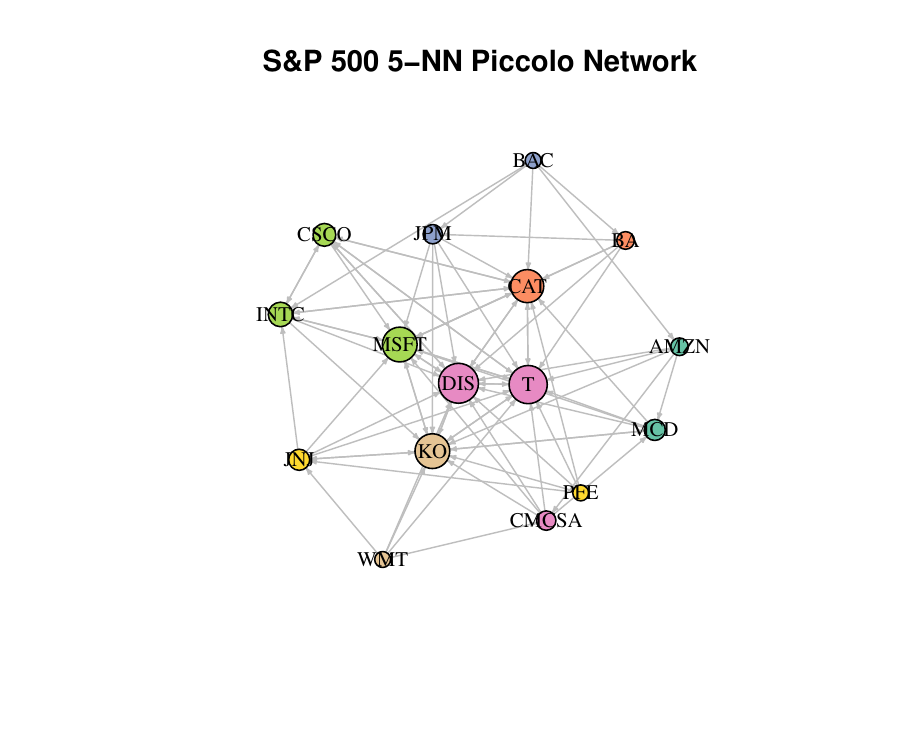}
      
    \end{minipage}

    \vspace{1em}
    \begin{minipage}[t]{0.31\linewidth}
        \centering
        \includegraphics[width=\linewidth]{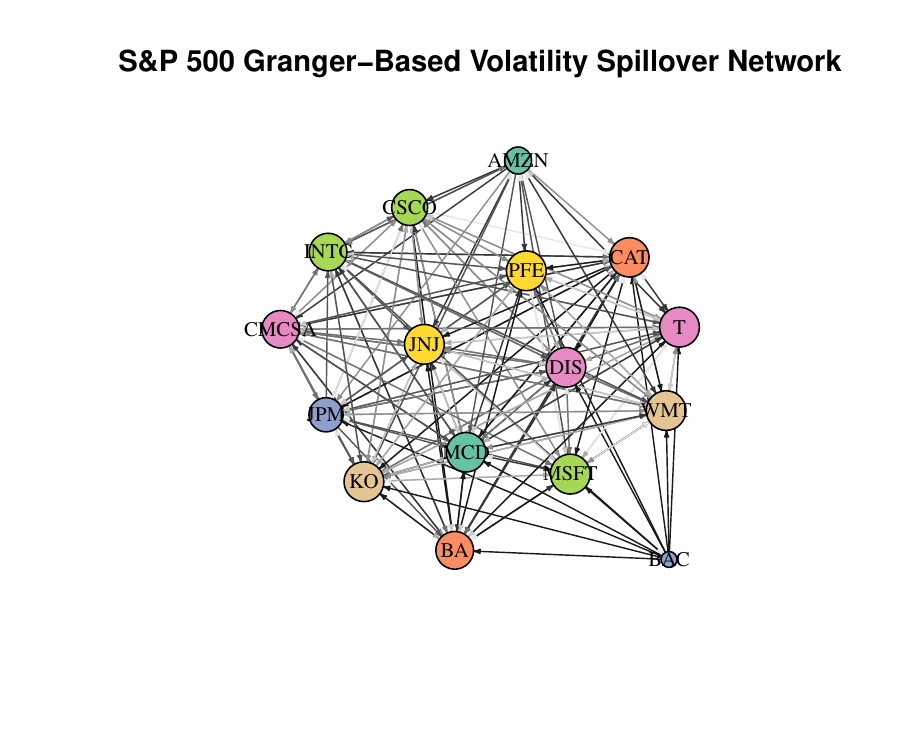}
    \end{minipage}%
    \hfill
    \begin{minipage}[t]{0.31\linewidth}
        \centering
        \includegraphics[width=\linewidth]{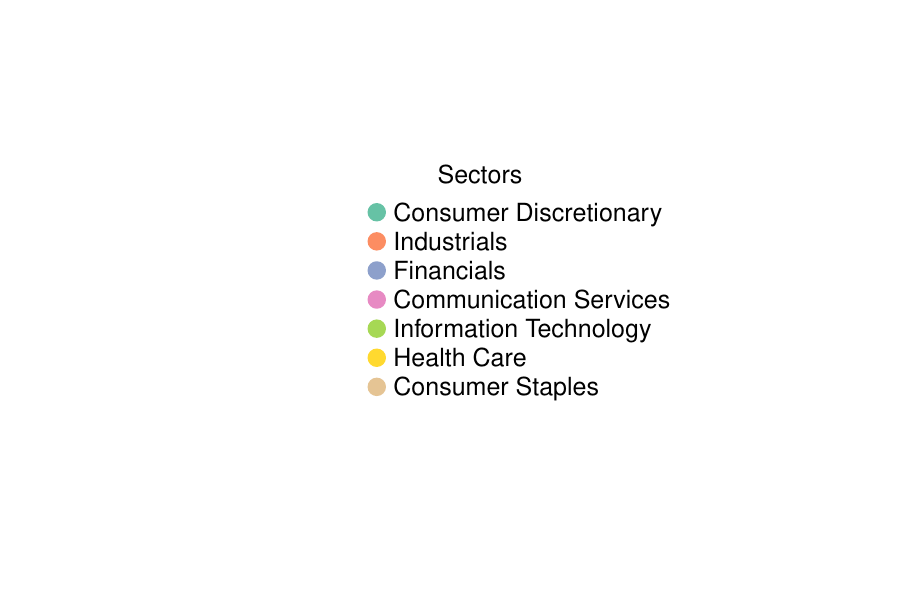}
    \end{minipage}

\end{figure}

\end{appendix}

\end{document}